%% file: Raft_Arxiv.tex
\documentclass{article}

\input{definitions}

\usepackage{arxiv}

\usepackage[utf8]{inputenc} % allow utf-8 input
\usepackage[T1]{fontenc}    % use 8-bit T1 fonts
\usepackage{hyperref}       % hyperlinks
\usepackage{url}            % simple URL typesetting
\usepackage{booktabs}       % professional-quality tables
\usepackage{amsfonts}       % blackboard math symbols
\usepackage{nicefrac}       % compact symbols for 1/2, etc.
\usepackage{microtype}      % microtypography
\usepackage{lipsum}
\usepackage[font=small,labelfont=bf]{caption} 
\usepackage{xcolor}

\usepackage{caption}
\usepackage{multirow}
\usepackage{gensymb}%per mettere degree
\usepackage[font=small,labelfont=bf]{caption} %per il carattere delle caption
\usepackage{float}

\usepackage{booktabs}
\usepackage{graphicx} 
\usepackage{siunitx} 

\usepackage[toc]{appendix}
\numberwithin{equation}{section}
\usepackage{pdflscape}

\usepackage{rotating}
\usepackage{tabularx}
\usepackage[labelfont=bf]{caption} % optional

\usepackage{graphicx,calc}
\usepackage{amsmath}
\usepackage[english]{babel}
\usepackage{eurosym}
\usepackage{stmaryrd}
\usepackage{caption}
\newcommand\ff{\textbf{\textit{f}}}

\everymath{\displaystyle}

\newcommand{\Fig}[1]{Figure \ref{#1}}
\newcommand{\Tab}[1]{Table \ref{#1}}
\newcommand{\sss}[1]{\scriptscriptstyle{#1}}

\graphicspath{{images//}}

\title{Mechanobiology predicts raft formations triggered by ligand-receptor activity across the cell membrane}

\author{
  A. R. Carotenuto\\
  Department of Structures for Engineering and Architecture, University of Napoli "Federico II" - Italy\\
  \texttt{angelorosario.carotenuto@unina.it} \\
  \And
  L. Lunghi \\
  Smiling International School,\, formerly at the Department of Life Sciences and Biotech., University of Ferrara - Italy\\
  \texttt{lnglra1@unife.it}\\
   \And
   V. Piccolo\\
   Department of Civil, Environmental and Mechanical Engineering, University of Trento - Italy\\
   \texttt{valentina.piccolo@unitn.it}\\
  \And
   M. Babaei\\
   Department of Civil and Environmental Engineering, Department of Mechanical Engineering, Carnegie Mellon - USA\\
   \texttt{mbabaei@andrew.cmu.edu}\\
   \And
   K. Dayal\\
   Department of Civil and Environmental Engineering, Department of Mechanical Engineering, Carnegie Mellon - USA\\
   \texttt{kaushik@cmu.edu}\\
   \And
   N. M. Pugno\\
   Department of Civil, Environmental and Mechanical Engineering, University of Trento - Italy;\\
   Laboratory of Bio-inspired, Bionic, Nano, Meta Materials \& Mechanics, Department of Civil,\\
   Environmental and Mechanical Engineering, University of Trento - Italy;\\
   School of Engineering and Materials Science, Queen Mary University of London, Mile End Road, London E1 4NS- UK\\
   \texttt{nicola.pugno@unitn.it}\\
   \And
   M. Zingales\\
   Dipartimento di Ingegneria, Università di Palermo, viale delle Scienze ed.8, 90128 Palermo, Italy\\
   \texttt{massimiliano.zingales@unipa.it}\\
   \And
   L. Deseri\\
   Department of Civil, Environmental and Mechanical Engineering, University of Trento - Italy;\\
   Department of Civil and Environmental Engineering, Department of Mechanical Engineering, Carnegie Mellon 	- USA;\\
   Department of Mechanical Engineering and Material Sciences, SSoE, University of Pittsburgh - USA; \\
   Department of Nanomedicine, The Houston Methodist Research Institute - USA\\
   \texttt{Lud7@pitt.edu}
   \And
  M. Fraldi\\
  Department of Structures for Engineering and Architecture, University of Napoli Federico II, Italy\\
  \texttt{fraldi@unina.it} 
}

\begin{document}
\maketitle

\begin{abstract}
Clustering of ligand-binding receptors of different types on thickened isles of the cell membrane, namely lipid rafts, is an experimentally observed phenomenon.  
Although its influence on cell's response is deeply investigated, the role of the coupling between mechanical processes and multiphysics involving the active receptors and the surrounding lipid membrane during ligand-binding has not yet been understood. 
Specifically, the focus of this work is on \textit{G-protein-coupled receptors} (GPCRs), the widest group of transmembrane proteins in animals, which regulate specific cell processes through chemical signalling pathways  involving a synergistic balance between the \textit{cyclic Adenosine Monophosphate} (cAMP) produced by active GPCRs in the intracellular environment and its efflux, mediated by the  \textit{Multidrug Resistance Proteins} (MRPs) transporters.
This paper develops a multiphysics approach based on the interplay among energetics, multiscale geometrical changes and mass balance of species, i.e. active GPCRs and MRPs, including diffusion and kinetics of binding and unbinding. Because the obtained energy depends upon both the kinematics and the changes of species densities, balance of mass and of linear momentum are coupled and govern the space-time evolution of the cell membrane. 
The mechanobiology involving remodelling and change of lipid ordering of the cell membrane allows to predict dynamics of transporters and active receptors --in full agreement with experimentally observed cAMP levels-- and how the latter trigger rafts formation and cluster on such sites. 
Within the current scientific debate on Severe Acute Respiratory Syndrome CoronaVirus 2 (SARS-CoV-2) and on the basis of the ascertained fact that lipid rafts often serve as an entry port for viruses, it is felt that approaches accounting for strong coupling among mechanobiological aspects could even turn helpful in better understanding membrane-mediated phenomena such as COVID-19 virus-cell interaction.
\end{abstract}

% keywords can be removed

\section{Introduction}\label{intro}

Cellular communication relies upon binding of ligands to specific cell surface receptors. \textit{G-protein-coupled receptors} (GPCRs) are key players in initiating and regulating cellular processes as they mediate responses to hormones, neurotransmitters, metabolites, ions, fatty acids, pathogens, and physical stimuli, such as light, smell, taste, and mechanical stretch. Moreover they can be activated by synthetic agonists, inhibited by antagonists and inverse agonists, or affected by allosteric modulators. They actually represent the most important superfamily of clinical targets in disorders of neural, immune, cardiovascular,  endocrine, respiratory system and cancer \cite{Cole:2011, Unal:2012, Allard:2012, Mary:2013}.
In particular, for disease like asthma, it is known to involve specific GPCRs, called $\beta_2$-\textit{adrenergic receptors}. This is because asthma entails a  continuous use of $\beta_2$-agonists, which are located in human airway epithelial cells. Indeed, this  results in loss of bronchoprotective effects and deterioration of asthma control.
The importance of such receptors in cancer relates to the fact that cAMP-based markers have shown to have a great potential for the early diagnosis of certain tumors. Groundwork for translation of the so called $\beta$-blockade as a novel adjuvant to existing therapeutic strategies in clinical oncology is on its way (see e.g. \cite{Cho:2000, Cvijic:2000, Liu:2010, Banerjee:2016, Choi:2018}).

\begin{figure}[htbp]
\centering
\includegraphics[width=0.7 \columnwidth]{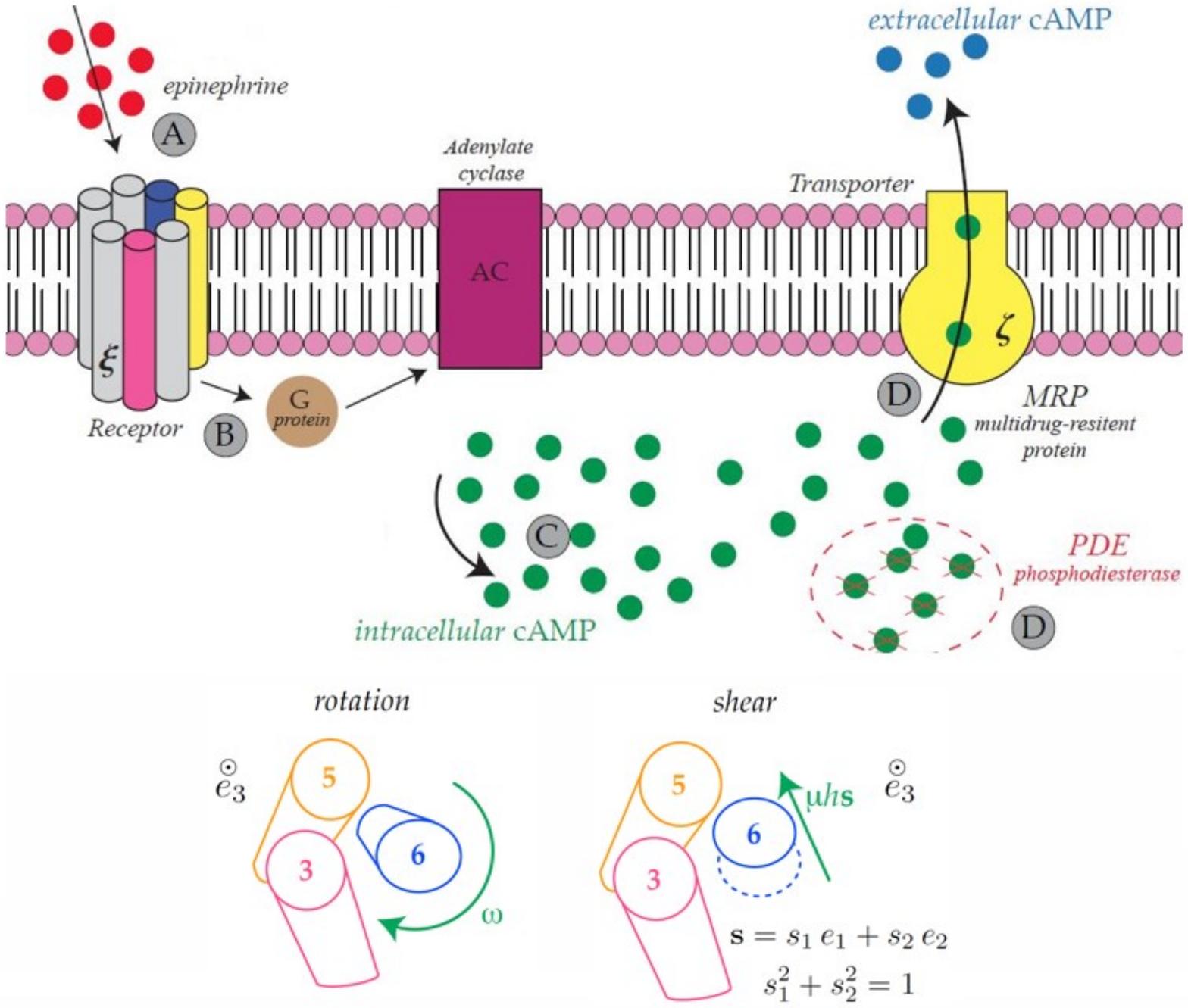}
\caption{Schematic representation of the ligand-binding between epinephrine (red dots on the top-left) and GPCRs (left, formed by seven cylindrical Transmembrane Domains-TMs) with production of intracellular and extracellular cyclic-Adenosine MonoPhosphate-cAMP- in the presence of transporters-MRPs (yellow unit to the right). Intermediate units are the G-protein (brown circle, near the GPCRS) and the Adenylate Cyclase (purple unit, across the membrane). The surface density of GPCRs bound to ligands is denoted by $\xi$, while the analog for MRPs is labelled as $\zeta$. The letters A, B, C and D display the order in which the chain of events occurs from the time in which ligand-binding takes place to the  effects of the cell response through the production of intracellular and extracellular cAMP.  The scheme highlights that a broad
spectrum inhibitor of cAMP phosphodiesterase, labelled as PDE, is accounted for \cite{Biondi:2010}. Helical domains forming the GPCR are displayed as cylinders in the figure: the TMs, where TM3, TM5 and TM6 have been reported in colors, to highlight their conformational changes. $\,$ The latter are displayed at the bottom of this figure. They consist in a rotation ($\omega$) of such domains about their axis, orthogonal to the (local) mid plane of the membrane (whose unit vector is denoted by $\bold{e}_3$), and on the shear ($\mu$) of TM6 approximately towards TM5, in the direction $\bold{s}$, whose components in the local mid-plane are also displayed in the figure (see e.g. \cite{Ghanouni:2001}).}
\label{fig_cell_with_receptors_and_transporters}
\end{figure}

GPCRs are major modulators of communication between the internal and external milieu of cells. These receptors are integral membrane proteins with an extracellular N-terminus and seven \textit{TransMembrane} helical \textit{domains} (TMs), from TM1 to TM7, connected by loop regions. Nowadays GPCR signaling is recognized to be more complex that was originally understood \cite{Kobilka:2007}. Briefly, their binding to very different kinds of extracellular stimuli induces TM domain \textit{conformational changes} and the structural remodelling of the protein. The latter allows for the coupling with cytoplasmic G proteins, which is followed by the activation of second messenger generating enzymes. The produced second messenger can activate many downstream signaling pathways inside the cell \cite{Lefkowitz:2007}. This activity of GPCRs is known to be detected through measurements of \textit{cyclyc Adenosyne Monophospate} -cAMP-, the detectable signature of the pathway arising in response to ligands, such as epinephrine. The production of cAMP in the cellular environment turns out to be modulated by several factors. In particular, the presence of \textit{Multidrug Resistance Proteins}-MRPs allows for the efflux of cAMP from the interior of the cell to the extracellular fluid, by so maintaining homeostatic intracellular concentrations. For this reason, MRPs are called transporters, as they contribute to keep a balance of cAMP inside the cell. A schematic of this process is depicted in Figure \ref{fig_cell_with_receptors_and_transporters}.

Historically, \textit{adrenergic receptors} are some of the most studied GPCRs. Two main subfamilies, $\alpha$ and $\beta$, which differ in tissue localization, ligand specificity, G protein coupling, and downstream effector mechanisms, are present. Diverse diseases, such as asthma, hypertension, and heart failure, are related to genetic modifications of adrenergic receptors. In particular, $\beta$2-adrenergic receptors ($\beta$2ARs) are found in smooth muscles throughout the body, and $\beta$2AR agonists are common treatments for asthma and preterm labor \cite{Cherezov:2007}. In \cite{Biondi:2006} it has been demonstrated the presence of $\beta$-adrenergic receptors in HTR-8/SVneo cell line, a well characterized first-trimester human extravillous trophoblast-derived cell line. Such cells invade the maternal uterine stroma, the decidua and the deeper portion of the myometrium evoking profound changes within the uterine vessels. A limited trophoblast invasion of maternal vessels has been correlated to both preeclampsia and fetal growth restriction, whereas an excessive trophoblast invasion is associated with invasive mole, placenta accreta and choriocarcinoma \cite{Lunghi:2007}. Other studies showed the importance of GPCRs during pregrancy \cite{LeBouteiller:2006, Hanna:2006}.

Ghanouni and coworkers  \cite{Ghanouni:2001} focused their studies on $\beta$2AR \textit{conformational changes} induced by agonist binding using quenching experiments. Their results can best be explained by either a rotation of TM6 and a tilting of TM6 toward TM5 during agonist-induced activation of the receptor. The structure, activation, and signaling of a GPCR is heavily influenced by the bilayer environment. This influence seems likely to involve either indirect bilayer effects, specific membrane-GPCR interactions, or a combination of both \cite{Drake:2006, Chini:2004, Oates:2011}. 

It has been reported that specific properties of this environment result in co-localization with downstream signaling components providing a rationale for regulation and specificity in GPCR activation \cite{Ostrom:2004, Patel:2008}. Some subsets of GPCRs are \textit{preferentially segregated to discrete regions} of the membrane defined as \textit{lipid rafts} 
\cite{Bray:1998, foster2003, Chini:2004, Ostrom:2004, Zhang:2005, Becher:2005, Barnett-Norris:2005, Gopalakrishnan:2005, Patel:2008, Watkins:2011, Villar:2016, Shukla:2016}. Despite lipid rafts are involved in a number of biologically relevant phenomena and their formation has been demonstrated to often serve as entry port for some viruses, including SARS-CoV ones \cite{lu2008lipid}, why is this the case and what is the physics at the basis of the associated remodelling of the lipid membrane %involving %the development functionally active such raft areas 
remain still open questions \cite{Gopalakrishnan:2005, Fallahi-Sichani:2009, Jacobson2007}.  Lipid raft are highly dynamic thickened areas on the lipid membrane with different dimensions and a complex heterogeneous composition. This is due to the fact that such rafts host a variety of transmembrane proteins responsible of compartmentalizing cellular processes to isolated districts of the cell.  These planar regions form localized microdomains on the cell membrane  %. The latter is characterized by a complex mixture of different kinds of lipids and active proteins triggering cellular processes 
and are hardly directly observable \textit{in vivo} \cite{risselada2008, DHARANI2015123}. \color{black}  In particular, they appear as dynamic nanoscale assemblies, where a high glycosphingolipid and cholesterol content in the outer leaflet of the lipid bilayer is present. This composition gives them a gel-like liquid-ordered organisation in comparison with the surrounding phospholipid-rich disordered membrane \cite{Chini:2004, Lingwood:2010}. 
%
%Indeed, the lipid structure is obviously well known, as it 
 
Each lipid molecule, composed by a hydrophilic head and a hydrophobic tail, may exhibit two different shape conditions of the tail, either straightened and taller (also known as \textit{ordered state}, $L_o$), or curly and shortened (also known as \textit{disordered state}, $L_d$). Such states depend on several conditions, among which the temperature and chemical composition of the lipid mixture are the main factors (see e.g. \cite{Bermudez:2004, Das:2008, Iglic:2012, Sackmann:1995}). 

The interplay between lipids and proteins is known to influence the overall behavior of the cell membrane. In particular,the adenosine A1, $\alpha$1-AR, $\beta$1-AR, $\beta$2-AR, AT1R, the endothelin (ETA-A and ET-B) receptors, and the M2-muscarinic receptors have all been localized to lipid rafts and/or caveolae \cite{Drake:2006}. 
The association of $\beta$2AR with caveolin3 in myocytes membranes has been demonstrated, as well as the fact that the confinement of $\beta$2-AR to caveolae is of critical importance for regulation of the intrinsic contraction rate in these membrane  preparations \cite{Xiang:2002}. How cholesterol modifies GPCR activity appears very much receptor-dependent, with regards to both upregulation and downregulation and of direct and indirect action \cite{Oates:2011, Hanson:2008}. 
The mechanical and chemical response of these amazing structures depends on many factors, such as the shape of the membrane, the temperature of the environment, the osmotic pressure, the chemical composition of lipid mixture, etc. \cite{Hu:2012, Norouzi:2006, Agrawal:2008, Agrawal:2009, Walani:2015, Baumgart:2003, Baumgart:2005,Honerkamp:2008}. 

Depending on the presence of embedded specialized proteins into the lipid membrane, several predicting models have been developed \cite{DeseriPZD:2015, Deseri:2013, Deseri:2008, Zurlo:2006, Canham:1970, Jenkins:1977, Agrawal:2009, Biscari:2002, Rangamani:2014, Walani:2015}. Most of them are based on the characterization of the energy, with the aim of predicting the ordered-disordered phenomena arising in transition regions for contrasting the external stimuli. Indeed, thickness reduction of the bilayer is observed by changing the mechanical and chemical conditions in the environment surrounding the cell. For this reason, several Authors adopted the change in thickness as order parameter for studying such kind of systems \cite{Goldstein:1989, Owicki:1979, Sackmann:1995}. Without explicitly doing so, a theory for the chemo-mechanical coupling of lipid bilayers accounting for curvature changes, viscosity, diffusion and $L_o-L_d$ phase transitions has recently been presented in \cite{Sauer:2017}. The $L_o-L_d$ phase transition has been modelled through a dimensionless order parameter related to the fraction of dioleoylphosphatidylcholine, known as DOPC. This is known to exhibit disordered features in agreement with the curliness of the tails mentioned above. Furthermore, binding and diffusion of certain proteins, epsin-1, is thoroughly analyzed. \color{black} As pointed out above, the physical behavior of such membranes is regulated by many factors enabling the occurrence of out-of-plane tractions, while in-plane shear stresses are not transmitted unless their viscosity is accounted for. 

The increasing availability of imaging techniques led indeed to a striking increase of  interest in the study of lipid membranes, often revealing  examples of the complex features characterizing their behavior (see, {\it e.g.}, \cite{Baumgart:2003, Baumgart:2005}). Bilayer elasticity has been fruitfully exploited in the literature  for the study of equilibrium shapes of red blood cells \cite{Jenkins:1977} and GUVs (Giant Unilamellar Vesicle) \cite{Canham:1970, Helfrich:1973}. The effects of embedded proteins or rod-like inclusions in the lipid membrane have been analyzed in \cite{Agrawal:2009, Walani:2015, Biscari:2002}, together with the analysis of buds formation \cite{Lipowsky:1992} with the coexistence of domains characterized by different bending rigidities \cite{Agrawal:2008, Baumgart:2005}.

Models for the exhibited order-disorder transition can be found in \cite{Akimov:2003, Chen:2001, Falkovitz:1982, Goldstein:1989, Iglic:2012, Jahnig:1981, Owicki:1978, Owicki:1979, Deserno:2014, Steigmann:2015} among others, while the effects of special  molecules (like cholesterol) on such transition have also been studied in \cite{Komura:2004, Pan:2009, Rawicz:2000, Akimov:2003, Honerkamp:2008, Lipowsky:1992}.

Models of lipid membranes where the bending behavior, the order-disorder transition and the chemical composition have been consistently taken into account can be found in \cite{Deseri:2008, Zurlo:2006, Maleki:2013, Agrawal:2009, Walani:2015, Steigmann:2013, Steigmann:2014a, Rangamani:2014, Deserno:2014}. In  \cite{Deseri:2008, Zurlo:2006}, the energetics regulating the behavior of such membranes was obtained through asymptotic dimension reduction (see also \cite{Deseri:2010}).
The major point in such papers is that the ``quasi-incompressibility'' of the environment, together with the influence of chemical composition, is enough to describe the membrane $L_o-L_d$ transition.

%=========== MANOSCRITTO NEW  NEW ==========
With the aim of understanding why GPCRs prefer to live on lipid rafts and, in turn, why the activation of transmembrane proteins involved in cell processes induce the formation of such thickened microdomains, the mechanobiology of the remodelling of the lipid bilayer is investigated by focusing on the interplay among membrane elasticity \color{black} and density changes of active species. In order to predict the dynamic remodelling of the cell membrane, following  thermodynamic arguments, a chemo-mechanical coupling is  found for the obtained chemical potentials of such species. These potentials regulate active species diffusion through mass balance which, in turn, involves the kinetics of binding through the occurence of interspecific terms. The potentials of the active species determine a corresponding coupling for the resulting energetics. Furthermore, it is found that the chemo-mechanical coupling found for such potentials is the specific work exterted by the lateral pressure (arising across the membrane thickness) against the volume changes around the interaction sites among the species domains and the surrounding lipids. It is worth noting that accounting for the aspects mentioned above enables mechanobiology to give a mechanically-based justification to the experimental findings of Nobel Prizes 2012 Kobilka \cite{Kobilka:2007} and Leifkovitz  \cite{Lefkowitz:2007}. \color{black}

%%%%%%%%%%%%%%%%%%%%%%%%%%%%%%%%
\section{Experimental measurements to trace protein fractions}
\label{ExpMeas}
%
%\renewcommand{\theequation}{A.\arabic{equation}}
  % redefine the command that creates the equation no.
\setcounter{equation}{0}  % reset counter
%%%%%%%%%%%%%%%%%%%%%%%%%%%%%%%%%%%%%%%%%%%
Experiments involving human trophoblast cells known to contain $\beta$-adrenergic receptors have been performed to the extent of detecting the activity of such GPCR in the human placenta (see e.g. \cite{Lunghi:2007}, \cite{Biondi:2010}).
The occurrence of a Multidrug Resistant Protein (MRP)-dependent cAMP efflux is also shown in human first-trimester placenta explants. Extracellular cAMP has been hypothesized to represent a source for adenosine formation that, in turn, could modulate cAMP-dependent responses in placental tissue. Evidence is provided that adenosine receptor subtypes are present and functional in \textit{Human TRophoblast} (HTR)-derived \textit{cells}. A role for cAMP egress mechanism in the fine modulation of the nucleotide homeostasis is then highly probable.
\subsection{Materials and Methods}
In the papers cited above, HTR-8/SVneo trophoblast cell line obtained from human first-trimester placenta ex-plant cultures and immortalized using SV40 large T antigen, donated by Dr. C.H. Graham, Queen$^{^,}$s University (Kingston, ON-Canada) to the Cell Biology Laboratory of the University of Ferrara, were cultured at 37$^\circ$C in a controlled atmosphere of $5\%$ $CO_2/95\%$ air in RPMI 1640 medium containing $10\%$ fetal bovine serum, 100 U/ml penicillin and 100 $\mu$g/ml streptomycin. These cells  were grown to confluence (2-3 days) in a 24-well plate (250,000 cells per well). The medium was then removed and replaced by serum-free RPMI.

The incubation process took place in wells of capacity $C_w=500 \,\mu l$ of such RPMI. This was carried out in the presence of isobutylmethylxanthine, otherwise known as IBMX, a broad
spectrum inhibitor of cAMP phosphodiesterase (PDE) \cite{Biondi:2010}. This was performed in the presence and in the absence of a \textit{given amount of epinephrine} for the indicated times. 

Media were then collected and immediately frozen at -70$^{\circ}$C until \textit{extracellular cAMP }(cAMP$_e$) levels were measured. Ice-cold 0.1 N HCl (0.25 ml) was added to the cells and, after centrifugation at 12,500 x g for 10 min, supernatants were neutralized adding 0.5 M Trizma base (0.05 ml) and utilized for measuring \textit{intracellular cAMP} (cAMP$_i$).
Finally, cAMP$_i$ and cAMP$_e$ were determined by standard methods (e.g. method of Brown \cite{Brown:1972}) and the nucleotide levels were expressed as $pmoles/10^6 cells/time$. %(\ref{fig:experiment}). 

%\begin{figure}[htb]
%\centering
%\includegraphics[width=0.98 \columnwidth]{figures/figG_experimental_setup.pdf}
%\caption{Schematic of the performed experiment: (a) trophoblast cell taken from human placenta; (b) harvesting in a 24wells plate; (c) treatment in controlled atmosphere; (d) cAMP measurement.}
%\label{fig:experimental_setup}
%\end{figure}

\begin{figure*}[t]
\includegraphics[width=0.45 \columnwidth]{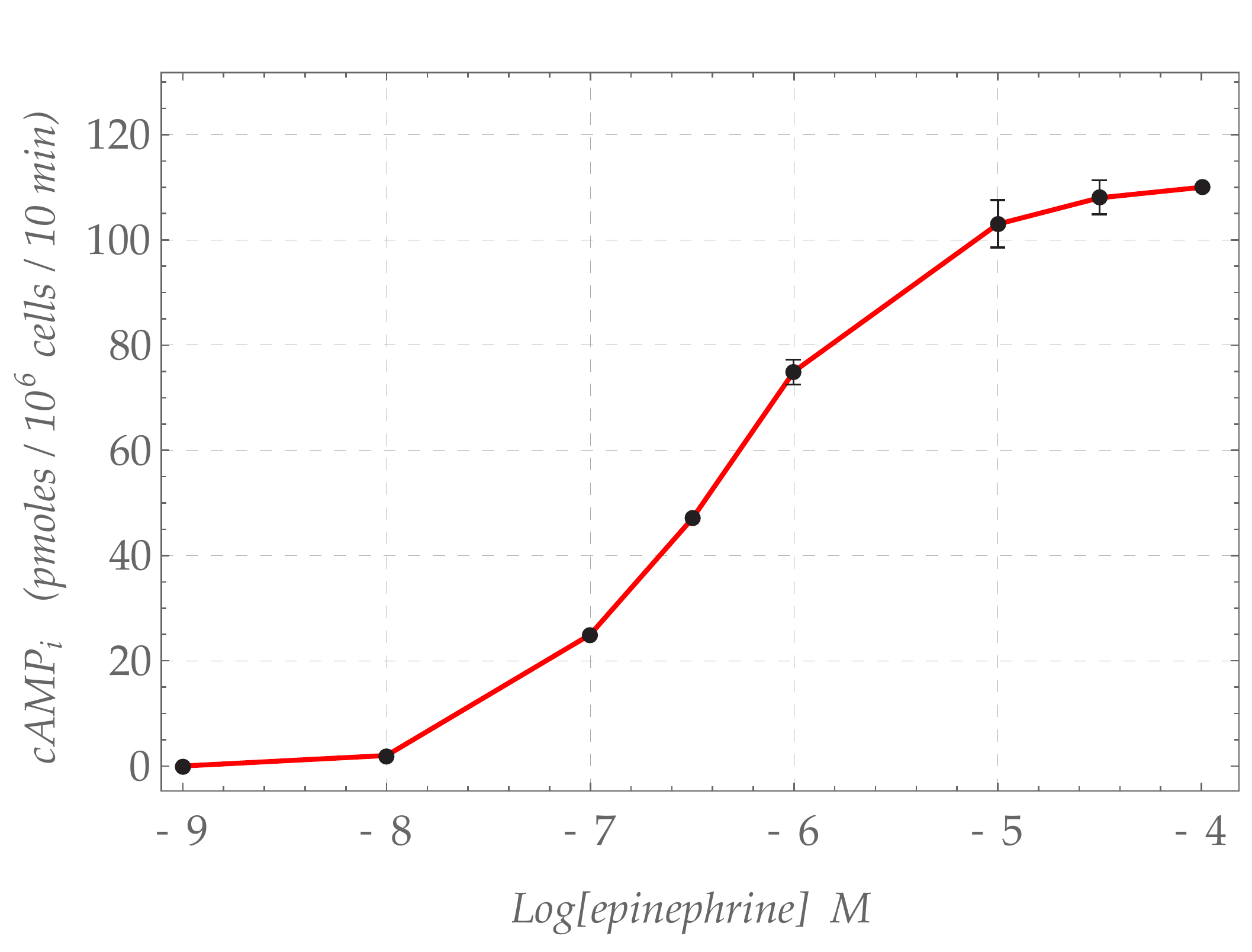} \quad
\includegraphics[width=0.45 \columnwidth]{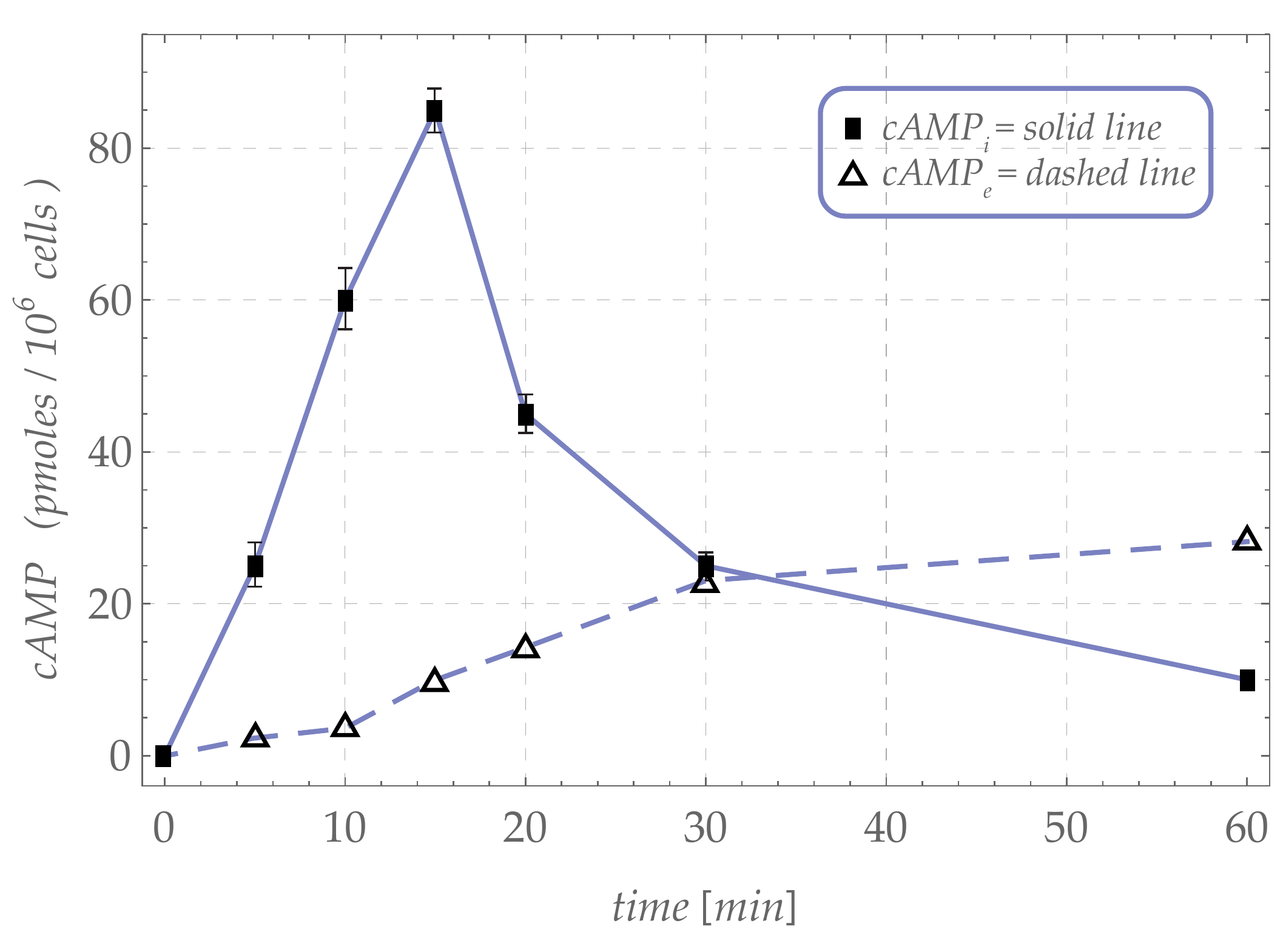} 
\caption{Experimental results: (a) value of measured intracellular cAMP after 10 minutes depending on epinephrine concentration, (b) time evolution  of measured intracellular cAMP for a fixed value of epinephrine concentration ($10^{-6} M$).}
\label{fig:experiment}
\end{figure*}

Tests with increasing concentration of epinephrine were first performed and cAMP measurements were recorded at a given time (10 minutes).
Epinephrine turned out to enhance intracellular cAMP in a dose-dependent fashion, reaching a plateau at around $10^{-4}\, M$ (Figure \ref{fig:experiment}\hyperref[{fig:experiment}]{A}). 

Furthermore, the cAMP evolution over time was monitored in the absense and in the presence of $10^{-6}\, M$ concentration of epinephrine. Such a concentration was chosen because it triggered a cAMP$_i$ production close to the half of the maximum response (Figure \ref{fig:experiment}\hyperref[{fig:experiment}]{B}). 

The cAMP levels were measured in cells incubated up to 60 minutes. 

In basal conditions, cAMP concentrations remained almost constant at all tested times (around $6.0$ pmoles/$10^6$ cells, not shown). In the presence of epinephrine, cAMP$_i$ increased as a function of the incubation time up to 15 min (14-fold), thereafter a reduction of the nucleotide level was observed. 

At the same time, cAMP$_e$ gradually increases in time, thereby almost reaching a plateaux, at least during the 60 min of observation (not shown). 

Analogous experimental observations in \cite{Biondi:2006, Lunghi:2007, Biondi:2010}, although with different ligands, showed an analogous trend. For the present case, when a concentration $c = 1 \,\mu M$ of epinephrine was used, the total quantity  $Q$ of ligand utilized in the experiment involving $10^6$ cells can be then computed as follows:
\begin{equation}
\label{eq:totalQ}
Q = 4 \times \, c \,C_w   =  4 \times 1 \, \mu M  \times 500 \, \mu L   =  2000 \,pmol .
\end{equation}

The experimental diagrams coming from the measurements of cAMP production are displayed in Figure \ref{fig:experiment}.

% ========================================
% cAMP-to-Receptors
% ========================================
\paragraph{cAMPi-to-$\xi$ and cAMPe-to-$\zeta$ relationships} The fields of active receptors density $\xi(x,t)$ and its products, namely the intracellular cAMP$_i$, are intimately linked by a conversion factor, namely $a_{\xi}$, as follows:
\begin{equation}
\label{eq:camptoxi}
cAMP_i = a_{\xi} \left(\xi - \bar{\xi} \right) \approx a_{\xi}\, \xi ,
\end{equation}
where $a_{\xi} = 10^4$ is estimated experimentally \cite{Biondi:2010} and $\bar{\xi}$ is the basal value of the density of active receptors. Here, the cAMP level is referred to a population of $10^6$ cells (and it is expressed in $pmol$). Taking into account a parameter $c_1 = 10^6$ for switching between the population and the single cell, and $c_2 = 10^{12}$ for converting $pmol$ to $mol$, the number of active receptors can be computed as follows:
\begin{equation}
\label{eq:cAMP-to-Xi}
\xi_{\#} = \frac{1}{c_1}\frac{1}{a_{\xi}} \left( \frac{cAMP_i}{c_2}\right) N_A ,
\end{equation}
where $N_A$ is the Avogadro number. An analogous relationship holds for computing the number of transporters:
\begin{equation}
\label{eq:camptozeta}
cAMP_e = a_{\zeta} \left(\zeta - \bar{\zeta} \right) \approx a_{\zeta} \zeta,
\end{equation}
where $\bar{\zeta}$ represents the basal value of the density of active transporters. In both the cases, direct proportionality between cAMP concentrations has been used to obtain an indirect estimation of the protein fractions predicted by the dynamics.
%

%%%%%%%%%%%%%%%%%%%%%%%%%%%%%%%%%%%%%%%%%%%%%%%%%%%%%%%%%%%%%%%%%%%
%\section*{Materials and Methods}
\section{The biomechanical model}
\label{sec:2}
The mechanobiology of lipid raft formation is investigated by accounting for deformation localization phenomena during binding of receptors, namely GPCRs compounds, with the incoming ligand. To this \color{black} aim, \color{black} the cell membrane is modelled as a thin hyperelastic soft body on which \textit{active receptors} with density $\xi$ and \textit{transporters} (i.e. MRPs-Multidrug Resistant Proteins) with density $\zeta$ mediate the communication between the extra-cellular and the intra-cellular environment. \color{black} Receptor-Ligand (RL) compounds and MRPs directly initiate different signaling pathways involved in different cell processes, regulated by the transport and compartmentation of chemicals such as cyclic-AMP.  Measuring cAMP concentrations then allows to indirectly estimate the amount of surface proteins involved, as  explained in Section \ref{ExpMeas}. There the experiments used to trace the levels of receptors and transporters are described.  In the proposed model, the cell membrane is characterized according to the constitutive framework proposed by Deseri and Zurlo in \cite{Deseri:2008, Deseri:2013, Zurlo:2006}. Here, this is explicitly coupled with the dynamics of GPCRs and MRPs activation, apt to describe the growth of active domains on the membrane surface by means of flow rules that include the receptor-transporter interplay. This dynamics is modelled according to a theory of interspecific growth mechanics recently proposed by Fraldi and Carotenuto in \cite{fraldi2018cells, carotenuto2018} and applied to model the growth of solid tumors. Here, the same framework is adapted to model the kinetics and diffusion of transmembrane proteins triggered by ligand-binding and mediated by the membrane elasticity. \color{black} Upon utilizing this approach, the evolution equations for the species provide Lotka-Volterra-inspired interspecific terms aimed to model the response of GPCRs to \color{black} affine ligands and the growth of RL complex. The latter stimulate the activation of MRPs through the intracellular cAMP production. In turn, it is observed a \color{black} localization of islands of active receptors and transporters on thickened domains of the cell membrane, known as lipid rafts. Such remodelling phenomenon is then governed by kinetics, diffusion and energetics, where binding/unbinding events turn out to be strongly coupled to the mechanical response of the plasma membrane through the energetics. Indeed, the mechanical interaction between the membrane and the active proteins during their remodelling and the conformational changes of their domains can be traced through the growth of their corresponding fractions. \color{black} \\
%
%
%
%==============================================
% ELASTICITY
%==============================================
\subsection{Membrane elasticity model for lipid bilayers}
\label{sectionB3}
In this Section the main results obtained in \cite{Deseri:2008, Deseri:2013, Zurlo:2006} are briefly recalled, and a schematic description of the approach followed in these papers is provided. The formulation of the model is based on the following assumptions: 
\begin{itemize}
\item[(i)] the effects leading to a spontaneous or natural curvature of the bilayer are ignored, i.e. it is assumed that the natural configuration is flat; 
\item[(ii)] the membrane kinematics is restricted to the class of normal preserving deformations, i.e. the normal vector $\mathbf{n}$ remains always normal to the mid-surface.
\end{itemize}
\begin{figure}[htbp]
\centering
\includegraphics[width=0.5 \columnwidth]{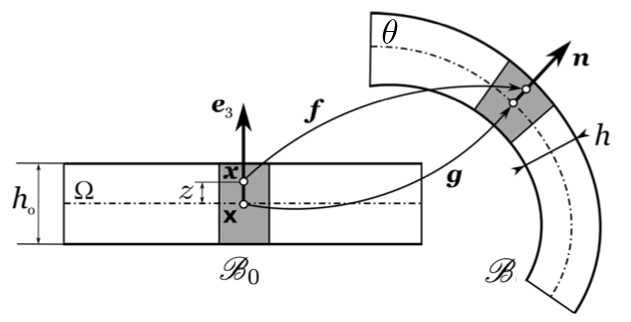}
\caption{Schematic representation of the deformation of a prismatic, plate-like reference configuration $\mathcal{B}_0$ into the current configuration $\mathcal{B}$. The gray box depicts the space occupied by two lipid molecules, their volume being conserved during the deformation. Courtesy of \cite{Deseri:2013}.}
\label{figB:fig00_schemagen}
\end{figure}

Under these assumptions, a simplified version of the elastic energy related to the change of the membrane geometry has been obtained in \cite{Deseri:2008, Deseri:2013}, The deformation mapping is denoted by $\mathbf{y}=\textbf{\textit{f}}(\textit{x})$, while its gradient is $\textbf{F}=\nabla\mathbf{y}= \nabla_\textit{x}\,\textbf{\textit{f}}$. The stored Helmholtz free-energy can be then expressed as
\begin{equation}
\mathcal{E}(\textbf{\textit{f}})=\int_{V^0}\Psi(\textbf{\textit{f}})\,dV^0=\int_{\Omega} \int_{-h_0/2}^{h_0/2}\Psi(\textbf{\textit{f}})\,dz\,d\Omega,
\end{equation}

where $\Psi$ denotes the Hemholtz energy density per unit of referential volume. It is easy to recognize that the surface density energy has the form:
\begin{equation}
\label{eqB:en2d}
\psi(\ff) = \int_{-h_0/2}^{h_0/2}\Psi(\ff)\,dz.
\end{equation}
%%%%%%%%%%%%
%%%
For a fixed temperature, the \textit{natural} configuration $\mathcal{B}_0$ of the membrane is assumed to coincide with a thin flat prismatic shape of homogeneous thickness $h_0$ in direction $\textbf{e}_3$, width $B$ in direction $\textbf{e}_2$ and length $L$ in direction $\textbf{e}_1$, with ordered  phase $L_o$ (see Figure \ref{figB:fig00_schemagen}). The membrane geometry is split into two entities, the two-dimensional mid-plane and the thickness, hence, the material particles $\textbf{\textit{x}}\,\in\mathcal{B}_0$ are described by:
\begin{equation}
\textbf{\textit{x}}=\textbf{x}+z\textbf{e}_3,
\end{equation}
where $\textbf{x}=x\,\textbf{e}_1+y\,\textbf{e}_2$  denotes locations of a flat mid-surface $\Omega$,  and $z$ spans the whole thickness. The reference membrane mid-surface $\theta$ corresponds to $z=0$, and its edges are defined by $x=\pm L/2$ and $y=\pm B/2$. Experimental evidences suggest that lipid membranes exhibit the so-called \textit{in-plane fluidity}, i.e. the absence of viscosity does not allow for sustaining shear stress in planes perpendicular to $\textbf{e}_3$. Based on this constitutive assumption, it is possibile to find a relationship for the energy density $\Psi$ depending on the three invariants of the deformation gradient, namely $\{\tilde{J}(\textbf{\textit{x}}),\,\det \textbf{F}(\textbf{\textit{x}}),\,\bar{\phi}(\textbf{\textit{x}})\}$.
%
%\begin{equation}
%\mathfrak{J}(\textbf{\textit{x}})=\{\tilde{J}(\textbf{\textit{x}}),\,%\det \textbf{F}(\textbf{\textit{x}}),\,\bar{\phi}(\textbf{\textit{x}})\},
%\end{equation}
%
In particular, the quantity $\tilde{J}(\textbf{\textit{x}}) := \sqrt{\det \, C_{\alpha \beta}}$ represents the areal stretch of planes perpendicular to the direction $\textbf{e}_3$, after setting $\textbf{C} = \textbf{F}^{T}\textbf{F }= C_{ij} \textbf{e}_i \otimes\textbf{e}_j$, which is the usual Cauchy-Green stretch tensor. The remaining invariants are the volume change, $\det\textbf{F}$, and the $\bar{\phi}(\textbf{\textit{x}}) = h(\textbf{\textit{x}})/h_0$ thickness stretch in direction $\textbf{e}_3$, respectively \cite{Deseri:2008, Deseri:2013}.
%%%
In \cite{Lipowsky:1992, Sackmann:1995} it is suggested that the volume of lipid membranes does not significantly change. Some authors showed  also that the volume of biological membranes can be assumed constant at several values of temperature \cite{Goldstein:1989, Owicki:1978}.  The following Ansatz (see Figure \ref{figB:fig00_schemagen}) is assumed for describing the geometrical changes of the membrane:
\begin{equation}
\label{eqB:def} 
\textbf{\textit{f}}(\textbf{\textit{x}}) = \textbf{\textit{g}}(\textbf{x})  + z\phi(\textbf{x}) \,\textbf{n}(\textbf{x}) 
\end{equation}
where \color{black} $\textbf{\textit{g}}(\circ)$ maps the mid-plane $\Omega$ of the membrane from the natural configuration $\mathcal{B}_0$ to the current mid-surface of the membrane (i.e $\theta=\textbf{\textit{g}}(\Omega)$), $\textbf{n}$ denotes the outward normal to $\theta$, and  $\phi(\textbf{x})=\bar{\phi}(\textbf{x})=h(\textbf{x})/h_0$ is the thickness stretch. The latter is defined as the ratio of the current thickness $h$ over the reference value $h_0$.  It is worth noting that a Monge representation $\bar{\textbf{\textit{g}}}(x,y):=\textbf{\textit{g}}(\textbf{x})$ for $\theta$ arises upon writing the position of material particles $\textbf{x} \in \Omega$ in components with respect to the pair $\textbf{e}_1, \, \textbf{e}_2$ of orthonormal vectors chosen above. The assumption of the Ansatz \eqref{eqB:def} is then coupled with a quasi-incompressibility constraint in the following form:
\begin{equation}
\label{eqB:quasinc}
\det\textbf{F}(\textbf{x},0)=\tilde{J}(\textbf{x},0) \phi(\textbf{x}) = 1.
\end{equation}
It is worth noting that such quasi-incompressibility is a first-oder approximation of the full iscochoricity requirement. An explicit expansion of equation \eqref{eqB:en2d} in powers of the reference thickness $h_0$ (see \cite{Zurlo:2006, Deseri:2008, Deseri:2013}) can be done by taking into account the choice of the constraint \eqref{eqB:quasinc}. Therefore, a restriction $\psi$ of the energy density $\Psi$ to $\Omega$ is considered by taking into account the requirement \eqref{eqB:quasinc}:
\begin{equation}
\label{eq:psiJ}
\psi(J)=\Psi(\tilde{J},\det\textbf{F},\bar{\phi}){\Big |}_{z=0}=\Psi(J,1,J^{-1}) ,
\end{equation}
where 
\begin{equation}
J \, \, :=\, \tilde{J}(\textbf{x},0).
\label{J_def}
\end{equation}
%%%%%%%%%%%%%%
Note that
\begin{equation}
\label{eq:J}
J=\left(\frac{h}{h_0}\right)^{-1}.
\end{equation}
In-plane fluidity, bulk incompressibility, and a dimension reduction \color{black} yield the following expression for the energy density of the lipid membrane $w$ per unit area in the reference configuration %(spatial descriptions of the quantities are indicated with the superscript $\widehat{\circ}$) \color{black}  
\cite{Deseri:2008, Deseri:2013, deseri:2012}
\begin{equation}
\label{eqB:en2dexp}
w = \varphi(J) + k(J) H^2 + k_G(J) K + \alpha(J) \, ||(\grad_{\theta} \widehat{J} )|_m||^2 ,
\end{equation}
\color{black} where $J$ accounts for the areal stretch of the membrane (see e.g. \cite{Wada:2015} for the importance of this quantity in lipid membranes), $H$ and $K$ are the mean and Gaussian curvatures of the mid-surface $\theta$, respectively, \color{black} $k(J)=\frac{h_0^2}{6}\tau'(J)$ and $k_G(J)=\frac{h_0^2}{12J} \tau(J)$ are the corresponding bending rigidities, $\varphi(J):= h_0 \, \psi(J)$ represents the \textit{local energy density} per unit area\color{black}, whereas 
\begin{equation}
\label{eqB:alpha}
\alpha(J)=\frac{h_0^2}{24}\frac{\tau(J)}{J^3},
\end{equation}
where
\begin{equation}
\label{eq:tau_def}
\tau(J) = \varphi'(J).
\end{equation}

The quantity $\alpha(J)$ is a higher order extensional modules and it is related to the nonlocal part of the energy density. In equation \eqref{eqB:en2dexp}, $\hat{J}$ is the spatial description of $J$ , defined by the composition $\hat{J} \circ g = J$, 
$\grad_{\theta}$ is the gradient with respect to points of the current mid-surface $\theta$, while $(\cdot)_{\scriptscriptstyle m}$ gives its material description.
The term \eqref{eqB:en2dexp} recovers exactly the Helfrich energy \cite{Helfrich:1973} if $J$ is constant. The main ingredient of the two-dimensional membrane model derived in \cite{Deseri:2008, Deseri:2013} is the surface Helmholtz energy $\varphi(J)$, which regulates the in-plane stretching behavior of the membrane and describes the phase transition phenomena taking place in lipid bilayers.
%%%%

% == FIGURE ==
\begin{figure}[htbp]
\centering
\includegraphics[width=0.5\columnwidth]{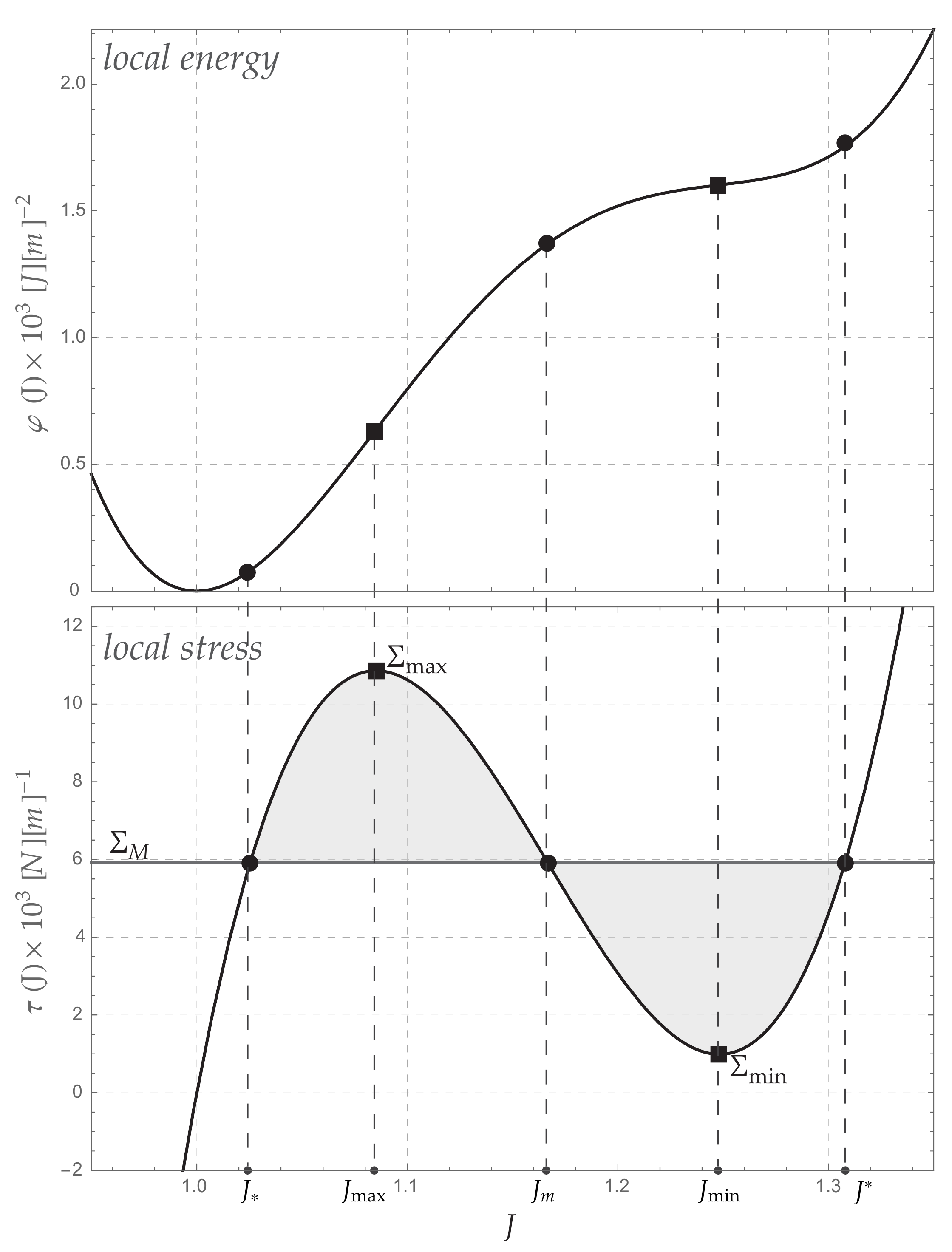} 
\caption{The stretching energy $\varphi(J)$ (top) adapted from \cite{Goldstein:1989} for a temperature $T\sim 30^{\circ}C$, with related local stress $\varphi'(J) = \tau(J)$ (bottom). The areal stretch $J_o=1$ corresponds to the unstressed reference configuration $\mathcal{B}_0$, while $\Sigma_M$ indicates the Maxwell stress.}
\label{figB:fig_dz_energy}
\end{figure}
% ======
A Landau expansion for  $\varphi(J)$ can be constructed because of the lack of more precise information (see, {\it e.g.}, \cite{Falkovitz:1982, Goldstein:1989, Komura:2004, Owicki:1978, Owicki:1979}). This approach results to be useful as it allows for relating the coefficients of the polynomial expansions to measurable quantities, such as the transition temperature, the latent heat and the order parameter jump (see \cite{Goldstein:1989} and the treatise \cite{Sackmann:1995} for a detailed discussion). 
Under the current assumptions on the natural configuration $\mathcal{B}_0$  the Landau expansion of stretching energy \cite{Zurlo:2006} takes the form:
\begin{equation}
\label{eqB:phi}
\varphi(J) = a_0 + a_1 J + a_2 J^2 + a_3 J^3 +a_4 J^4,
\end{equation}
where the parameters $a_i\,\,(i=0 \div 4)$ depend on temperature and chemical composition. Because of the lack of avalaible experimental data, these parameters have been calibrated in \cite{Zurlo:2006} by considering experimental estimates provided by \cite{Goldstein:1989, Komura:2004, Komura:2006}. For a temperature $T\sim 30^{\circ}C$ we have:
\begin{equation}
\label{eqB:ai}
\begin{array}{c}
a_0=2.03 , \quad a_1=-7.1, \quad a_2=9.23 \\ 
a_3=-5.3, \quad a_4=1.13
\end{array} 
\end{equation}
expressed in $[J][m]^{-2}$. The related local constitutive stress $\Sigma$ \eqref{eq:tau_def} shows the typical S-shaped form, as shown in Figure \ref{figB:fig_dz_energy}. The locations where the maximum $\Sigma_{\max}$ and the minimum $\Sigma_{\min}$ of the local stress occur are denoted by $J_{\max}$ and $J_{\min}$, respectively, and highlighted with squares, while the value of the Maxwell stress (determined by the  equal area rule) is drawn as a straight horizontal dark-grey line (see e.g. \cite{Coleman:1988}). Whenever a generic stress $\Sigma$ is considered, it is also possible to identify three intersections between such a stress and the local stress curve. In the case of the Maxwell stress, these three intersections are denoted with black circles, and the corresponding areal stretch values are called $J_*$, $J_m$ and $J^*$ from left to right. As an example, Table \ref{tabB:coleman} collects the numerical values of these quantities for $\varphi(J)$ obtained by using the coefficients in \eqref{eqB:ai}.
\begin{table}[htbp]
\centering
\caption{\textbf{Table}\ref{tabB:coleman}. Characteristic values of the membrane stretching energy at $T\sim 30^{\circ}$. Stress expressed as  $[\Sigma] =  [J/m^2] \, \times 10^{-3}$}
\label{tabB:coleman}
\begin{tabular}{cccccccc}
\hline
$\Sigma_{M} $  & $\Sigma_{\max} $ & $\Sigma_{\min} $  & $J_*$ & $J_m$ & $J^*$ &$J_{\max}$ & $J_{\min}$  \\
\hline
5.922 & 10.855 & 0.989 & 1.025& 1.167& 1.308& 1.085 & 1.248  \\
\hline
\end{tabular}
\end{table}

\subsubsection{Planar case}\label{sec:planarcase}
The study of the equilibrium for a planar lipid membrane described by the energy \eqref{eqB:en2dexp} permits to elucidate the emergence of thickness inhomogeneities in the membrane. Moreover, this simple energetics allows one to calculate the corresponding rigidities and the shape of the boundary layer between the ordered and disordered phases. By fixing temperature, energy density coefficients are fixed to those ones in Table \ref{tabB:coleman}. % in a way to also set $\widehat{\varphi}=\left.\varphi(J)\right|_{T=30^{\circ}C}$ from now on.
Whenever no curvature changes are experienced by the lipid bilayer, in the plane strain approximation the elastic energy density in \eqref{eqB:en2dexp} takes the form:
\begin{equation}
\label{eqB:energy_gamma_1D}
\psi(J) = \varphi\left(J\right) - \frac{1}{2} \gamma(J) J_x^2.
\end{equation}
We note that the functions 
\begin{equation}
\label{eqB:gamma}
\gamma(J) = -2\,\alpha(J).
\end{equation}
and $\tau(J)$, defined by \eqref{eq:tau_def}, can be interpreted as \textit{transition} and \textit{stress-like} functions, respectively, as the former drives the boundary layer wherever transitions between two phases occur. According to the geometry introduced above, the three-dimensional membrane deformation is further restricted with respect to equation \eqref{eqB:def} as follow:
\begin{equation}
\label{eqB:def2}
\textbf{\textit{f}}(\textbf{\textit{x}})=g(x)\textbf{e}_1+y\textbf{e}_2+z\phi(x)\textbf{e}_3 ,
\end{equation}
so that the width $B$ is kept constant and its gradient takes the following form
\begin{equation}
\label{eqB:F}
\textbf{F}=\nabla\textbf{\textit{f}}(\textbf{\textit{x}})=
  \left[
   \begin{array}{ccc}
      g_{x} & 0 & 0 \\
      0 & 1 & 0 \\
      z\phi_{x} & 0 & \phi
   \end{array}
   \right],
\end{equation}
where the subscript $x$ denotes differentiation with respect to $x$. The displacement component along $\textbf{e}_1$ is $u(x)=g(x) - x$. After setting
\begin{equation}
\label{eqB:lambda}
\lambda(x)= g_{x}(x),
\end{equation}
where the function $g$ maps the mid plane of the membrane from $\mathcal{B}_0$ to $\mathcal{B}$, see Figure  \ref{figB:fig00_schemagen}.
The plane strain approximation yields 
\begin{equation}\label{ThicknessChange}
\phi=\lambda^{-1}=\frac{h}{h_0},
\end{equation}
so that the membrane deformation is completely determined by 
\begin{equation}\label{J=lambda}
J = \lambda. 
\end{equation}

%%%%%%%%%%%%%%%%%%%%%%%%%%%%%%%%%%%%%%%%
%%%%%%%%%%%%%%%%%%%%%%
%                  REMODELING                     %
%%%%%%%%%%%%%%%%%%%%%%

\subsection{Membrane remodelling due to GPCR growth via densification} 
\begin{figure}[htbp]
\centering
\includegraphics[width=0.99\columnwidth]{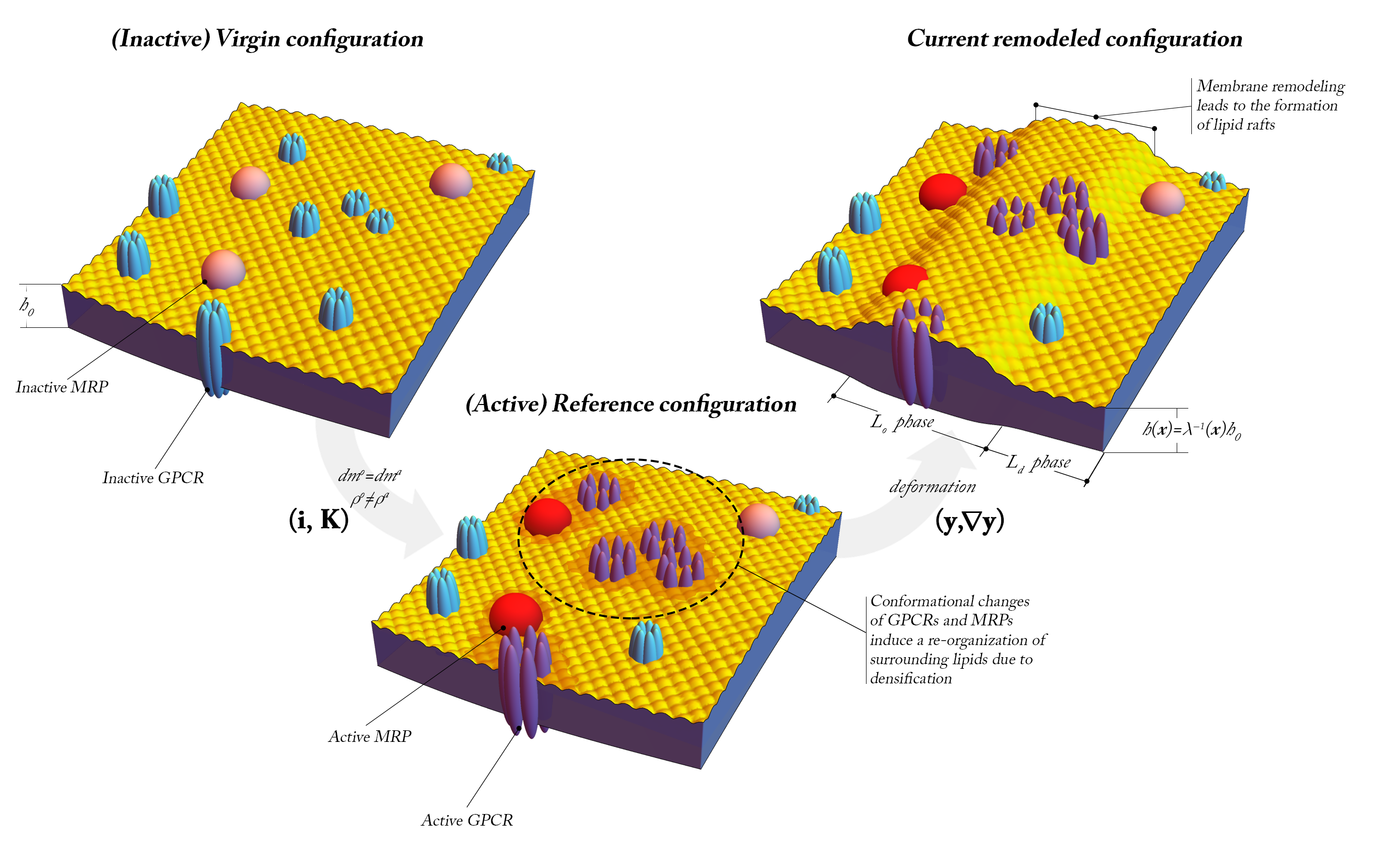} 
\caption{Schematic picture of protein activation and lipid membrane remodelling. A density change of active receptors drives the system towards an active configuration in which the lipid bilayer shrinks around active domains, this causing membrane deformation and thickening in the form of lipid rafts. Unlike usual description of finite deformations and multiphysics, all the three configurations are global. This is justified through a non-standard multiscale geometric framework provided by the theory of \textit{Structured Deformations} (see e.g.  \cite{deseri:2003}, \cite{OwenMix:2008}, \cite{Deseri:2010}, \cite{deseri:2015}, \cite{deseri:2019}, \cite{palumbo:2018}). Here, each material particle of the body in the virgin configuration (displayed in the top-left side of this figure) gets first mapped into an intermediate region (center of the figure) through a pair of smooth bijections $(\bf{i}, \bf{K})$ (where $\bf{i}$ stands for the vector-valued identity and the tensor $\bf{K}$ accounts for all the submacroscopic changes). Indeed, it can be shown that $K_r = det \bf{K}$ and, hence, this invariant of $\bf{K}$ accounts for remodelling. Secondly, each neighborhood of each particle located in this intermediate global configuration gets classically deformed through a pair $(\bf{y}, \nabla\bf{y})$, where $\bf{y}$ is a smooth deformation field from the intermediate global configuration onto the current one, while the second item in the pair is the deformation gradient from the intermediate global configuration. Features of Structured Deformations in connection with the examined biological systems are discussed in more details in  \hyperref[{SB2}]{Appendix C}. There, an alternative scheme of the kinematics is given in Fig.\ref{fig:StructuredDeformations}. 
}
\label{fig:Stress00}
\end{figure}

\color{black}  A thermodynamically consistent framework is provided in this section with the aim of evaluating the chemo-mechanical coupling between transmembrane proteins and membrane elasticity. To this end, a total Helmoltz free energy per unit volume is introduced, say $W^*$, accounting for both the response of the lipid bilayer and the energetics associated to the proteins changes. In particular, the rearrangement %activation
 of protein domains on the cell membrane is modelled as a process of growth via pure densification \cite{Lubarda}, i.e. by considering the effect of the variation of the density of activated proteins on the free-energy of the membrane. In fact, by considering a membrane element  with  mass $dm^0=\rho^0 dV^0$ in the inactive (virgin) configuration, as displayed in figure \ref{fig:Stress00}, the  potential  activation of previously dormant receptors on the membrane is likely assumed not to produce mass variation. 
However, the submacroscopic changes following the formation of receptor-ligand binding induce remodelling  and conformational changes of the bonded receptors across the membrane. This is because the variation of the density and of the conformations of those receptors determine the re-organization of the surrounding lipids (see the \hyperref[{SB2}]{Appendix C}).
By denoting with the superscript $a$ the active (reference) configuration in which protein activation occurs (see the inermediate global configuration of the membrane in figure \ref{fig:Stress00}), the remodelling $K_r$ affecting the mass element can be written as follows (see the \hyperref[{KrCalc}]{Appendix A}): \color{black}
\begin{equation}\label{RemodelingTerm}
\rho^0\, dV^0=\rho^a \,dV^a \rightarrow \,  
K_r := \frac{dV^a}{dV^0}=\frac{\rho^0}{\rho^a}=\frac{1+\kappa_u(1+\small{\Sigma}_i\,n_i^0\,\Delta_A)(\rho_p/\rho_l -1)}{1+\kappa_u(1+\small{\Sigma}_i\,n_i\,\Delta_A)(\varrho_p/\varrho_l -1)}.
\end{equation}
This factor is calculated by considering that the density of the heterogeneous medium in a certain configuration is the sum of the true densities weighted by the respective fractions, i.e. $\rho=\varrho_l\phi_l+\varrho_i\,n_i$ (see \cite{fraldi2018cells}), $\{n_i\}|_{\{i=1, \, 2\}}=\{\zeta, \xi\}$ representing the relative fractions of protein species (summation over \textit{i}) and $\phi_l$ the complementary lipid fraction, the additional superscript $0$ being used for indicating the initial ones. The other constants read \color{black} as  follows \color{black}:

\begin{itemize}
\item $\kappa_u=N_{max} A_u/A$, i.e. the total areal fraction of inactive proteins ($N_{max}=1.13\times 10^7$ is the total number of potentially activating proteins, estimated from the experimental data, $A_u$ is the area of the inactive domain (with diameter of $\approx 4 nm$ \cite{gurevich2018gpcrs}, $A$ is the surface of the cell for which a radius of $30 \mu m$ was considered)
\item $\Delta_A= (A_a/A_u-1)$ is the relative change in area of protein domains passing from inactive to active state, the sight of the latter exhibiting a diameter of $\approx 5 nm$ \cite{gurevich2018gpcrs}
\item the terms $\varrho_p$ and $\varrho_l$ indicate the specific true density of the transmembrane proteins and of the bilayer lipids, respectively, both assumed constants and calculated from their molecular weights and sizes. In particular, $\varrho_p=100 kDa/h_0 \pi (2.5 nm)^2$ and $\varrho_l=2\times(1 kDa/h_0 \pi (0.5 nm)^2)$ \cite{cevc1993, carpenter2016, mendelson2013, HUME2014363}.
\end{itemize}
%%%%
%\color{brown} It is worth notning that $K_r$ has a meaningful multiscale geometric interpretation through the \textit{Theory of Structured Deformations}, as highlighted in Fig.\ref{fig:Stress0}  (see e.g. \cite{deseri:2003}, \cite{Deseri:2010}, \cite{deseri:2012}, \cite{deseri:2015}, \cite{deseri:2019}, \cite{palumbo:2018}), as discussed in details in \hyperref[{SB2}]{Appendix C}.\color{black}

%\color{green} A schematic highlighting lipid membrane remodelling, as well as protein activation and geometrical changes involved in the membrane response during ligand binding is depicted in figure \ref{fig:Stress00} below.

% == FIGURE ==
% ====== 

\bigskip

\color{black} In the light of \eqref{RemodelingTerm}, the %remodelling-induced stretching
 energy stored through remodelling, stretching and species redistribution can be expressed in terms of the total (three-dimensional) Helmoltz free energy $W^*$ of the membrane in the natural configuration (i.e. the reference configuration in  fig. \ref{fig:Stress00}) as:
\begin{equation}\label{StretchingEnergy}
\int_{V^a} W^* dV^a=\int_{V^0}  K_r\,W^* dV^0=\int_{V^0}  W^*_0 dV^0
\end{equation}
%%%%%
where $W^*_0$ is instead the three-dimensional energy density relative to the %reference
virgin configuration.  In analogy with equation \eqref{eqB:en2d}, because of the thinnes of the cell membrane relative to its in-plane sizes, a dimension reduction is  performed to obtain a corresponding effective energy density $W_0$ per unit area in the virgin configuration:
\begin{equation}\label{position}
W_0=K_r\,W(\lambda,\nabla\lambda,n_i)=\int_{-h_0/2}^{h_0/2}W^*_0 dz=h_0\,K_r\,W^*(\lambda,\nabla\lambda,n_i).
\end{equation}

\subsubsection{Mass conservation equations and interspecific terms}
%Under isochoric assumptions, t 
The one-dimensional balance of mass for each constituents can be written as 
\begin{equation}\label{massbalance}
 \dot{n}_i+M_{i,x}=\Gamma_i
\end{equation}

where the dot notation denotes the material time derivative, $M_i$ describes the material species %transport 
flux and $\Gamma_i$ is the rate term describing the evolutionary behavior of each species. This is done by taking into account intrinsic rates as well as mutual interactions with other fields characterizing the system by means of suitable coupling terms. The present model mainly focuses on the active receptors and transport proteins on the cell membrane. In particular, in the presence of a specific %activator
ligand, binding and activation GPCRs kindles cytoplasmatic signaling pathways. The latter allow for both regulating some intracellular processes and communicating with the extra-cellular environment by expelling molecules %outside the cell 
through the activation of specific transport proteins, such as MRPs. %For example, t
This coupled mechanism can be followed in practice by measuring the levels of intracellular and extracellular concentrations of cAMP, involved in many physiological processes of the cell. In particular, intracellular cAMP production follows the response of the GPCRs to certain ligands, such as epinephrine. The level of cytoplasmatic cAMP is modulated by MRPs that permit the efflux of cAMP from the interior of the cell to the extracellular fluid. In this way, MRPs control cell homeostasis and determine an agonist interaction with membrane receptors.
In order to trace the interplay of the transmembrane proteins, we introduce mass balances for two selected species,  $n_1=\xi$, representing the G proteins, and  $n_2=\zeta$, namely the transport proteins. The main physiological aspects associated to their activity have been described by means of the following interspecific (Volterra-Lotka like) equations:
\begin{align}
 &\dot{\xi}+M_{\xi,x} = \alpha_\xi \, \xi -\delta_\xi \, \xi -\beta_{\xi\zeta} \, \zeta  \, \xi \label{massbalance2.1}\\
 &\dot{\zeta}+M_{\zeta,x} = \beta_{\zeta\xi} \, \xi \, \zeta -\delta_\zeta \, \zeta . \label{massbalance2.2}
\end{align}

\paragraph{Physical meaning of the species rates parameters.} In equation \eqref{massbalance2.1}, the term $\alpha_\xi$ denotes \textit{intrinsic activation in response to a given ligand concentration} 
(such as the epinephrine), modelled by means of a specific uptake function. In particular, it is likely assumed that the ligand precipitation rate obeys a generic Gamma distribution with probability density function $\Upsilon(t)=a\,t\,e^{-bt}$. The coefficients $a$ and $b$ are calibrated by means of experiments. By observing the time at which $\Upsilon$ is maximum -- i.e. $\dot{\Upsilon}(t_m)=0$ -- it readily follows that $b=t_m^{-1}$. Also, the total quantity of ligand precipitated over the experiment time $T$ then will be 
\begin{equation}\label{Q}
Q=\int_0^T \Upsilon(t) dt= ab^{-2} (1- e^{-bT}(1+bT))
\end{equation}
that allows to estimate the parameter $a={Q}/(t_m^2-e^{-\frac{T}{t_m}}t_m\left(t_m+T\right))$. Given this, the coefficient $\alpha_\xi$ results the relative uptake function multiplied by a suitable binding constant:
\begin{equation}\label{alphaxi}
\alpha_\xi=k_b \, Q^{-1}\Upsilon(T).
\end{equation}

The coefficient $\delta_\xi$ denotes an \textit{intrinsic deactivation constant}, while the \textit{interspecific term} $\beta_{\xi \zeta}$ is introduced in the light that the efflux of extracellular cyclic Adenosyne Monophospate, $cAMP_e$ (whose amount is proportional to the active MRP on lipidic membrane through \eqref{eq:camptozeta}), is assumed to reduce the activity of receptor-ligand compounds by establishing chemical equilibrium between the extra- and the intra-cellular space. Dually, in equation \eqref{massbalance2.2}, the rate of activation of transport proteins is stimulated by the level of intracellular cyclic Adenosyne Monophospate, $cAMP_i$ (which is proportional to the activated G-proteins at time $t$ through \eqref{eq:camptoxi}). This is accounted for through the \textit{interspecific positive term} $\beta_{\zeta\xi}$ governing the growth of activating MRPs. Model parameters have been therefore estimated through the experimental evaluation of intracellular and extracellular cAMP concentrations (see \hyperref[{ExpMeas}]{Section 2}).

\subsection{The Helmoltz free-energy of the membrane proteins}
\label{sec:3P}
The %total mechanical 
energy $W^*$ accounts for the hyperelastic contribution of the lipid membrane and for the Helmoltz free energy of the active proteins. The latter interact with the membrane by exchanging mechanical forces and showing remodelling changes during their activity. We start from the isothermal dissipation inequality. Here, we single out the contribution of mass transport of the %$i^{th}$ 
species over the generic elementary volume, say $M_i$ for the $i^{th}$ species, through the action of the correspoinding (thermodynamically conjugate) chemical potential $\mu_i^*$, as well as the presence of thermodynamic driving forces $F_r^*$ expending power against the rate of change of remodelling (see e.g. \cite{Lubarda, BMMB}). Hence, for an incompressible material and under isothermal assumptions, combined energy-entropy equations yield the following form of the second law of thermodynamics: 

\begin{equation}\label{dissip}
\int_{V^0}{P^*\dot{\lambda}}\,dV^0 - \int_{V^0}\left(\mu^*_i\,M_i\right)_{x}\,dV^0+\int_{V^0}F_r^* \dot{K}_r\,dV^0 \geq \frac{d}{dt}\int_{V^0}W^*_0  \, dV^0 
\end{equation}

\color{black} where repeated index $i$ means summation over that index. By appealing to relation \eqref{StretchingEnergy}, in agreement with the dimension reduction performed in \eqref{position}, quantities $X$ per unit area can be introduced by considering that $\int_{V^0}X^*dV^0=\int_\Omega X\, d\Omega$, with $X=h_0 X^*$ for any field $X^*$ per unit volume in the virgin configuration. 

\noindent Provided that 
\begin{equation}
\label{energy-reduced-reference}
W=W(\lambda,\lambda_x,n_i)
\end{equation}
from \eqref{position}, one also obtains the dimensionally reduced version of \eqref{dissip}:
\begin{equation}\label{dissip2}
\resizebox{.9 \textwidth}{!} {$
\int_{\Omega}{P\dot{\lambda}}\,dV^0 - \int_{\Omega}\left(\mu_i\,M_i\right)_{x}\,d\Omega+\int_{\Omega}F_r \dot{K}_r\,d\Omega \geq \int_{\Omega}\left[K_r\left(\frac{\partial W}{\partial \lambda}\dot{\lambda}+\frac{\partial W}{\partial \lambda_x}\dot{\lambda}_x +\frac{\partial W}{\partial n_i}\dot{n}_i\right)+W\,\dot{K}_r\right]  \, d\Omega .
$}
\end{equation}
By substituting the mass balance \eqref{massbalance} in the second integral of the left-hand side and by integrating by parts the second term on the right-hand side, equation \eqref{dissip2} can be rearranged as follows:
\begin{equation}\label{dissip3}
\resizebox{.9 \textwidth}{!} {$
\int_{\Omega}{\left[P\dot{\lambda}-\mu_{i,x}M_i +\mu_i\,\dot{n}_i -\mu_i\,\Gamma_i +F_r\dot{K}_r\right]}\,d\Omega 
\geq \left[\frac{\partial W_0}{\partial \lambda_x}\dot{\lambda}\right]_{\partial \Omega}+\int_{\Omega}\left\{K_r\left[\left(\frac{\partial W}{\partial \lambda}-\frac{\partial}{\partial x}\left(\frac{\partial W}{\partial \lambda_x}\right)\right)\dot{\lambda}+\frac{\partial W}{\partial n_i}\dot{n}_i\right]+W\,\dot{K}_r\right\}  \, dV^0 
$}
\end{equation}

Provided the condition
\begin{equation}\label{conditionstress}
\left[\frac{\partial W}{\partial \lambda_x}\dot{\lambda}\right]_{\partial \Omega}= 0
\end{equation}
is satisfied, grouping and localization of the terms in \eqref{dissip3} yields the following inequality 
\begin{align}\label{localdissip}
\left[P-K_r\left(\frac{\partial W}{\partial \lambda}-\frac{\partial}{\partial x}\frac{\partial W}{\partial \lambda_x}\right)\right]\dot{\lambda} +\left[\mu_i - K_r \frac{\partial W}{\partial n_i}\right]\dot{n}_i  + \left[F_r-W\right]\dot{K}_r -\mu_{i,x}M_i -\mu_i\,\Gamma_i \, \geq \, 0.
\end{align}

Following the standard Coleman and Noll's procedure, it is then possible to obtain the following constitutive relations for the Piola-Kirchhoff stress, the chemical potential and the remodelling driving force, which turns out to be entirely identifiable with the mechanical energy involved in the process:
\begin{align}\label{ConstEqual}
&P=K_r\left(\frac{\partial W}{\partial \lambda}-\frac{\partial}{\partial x}\frac{\partial W}{\partial \lambda_x}\right), \nonumber\\
&\mu_i=K_r\frac{\partial W}{\partial n_i},\\
&F_r=W. \nonumber
\end{align}

Inequality \eqref{localdissip2} thus reduces to:
\begin{equation}\label{localdissip2}
 -\mu_{i,x}M_i -\mu_i\,\Gamma_i\geq 0
\end{equation}

that can be fulfilled by expressing proteins movement term as proportional to the gradient of the chemical potential, i.e. $M_i =-L_i\,n_i\,\mu_{i,x}$, where $L_i\geq0$ are suitable mobilities. 
% Upon imposing 
A reduced dissipation inequality in the form $\mu_i\,\Gamma_i \leq 0$ remains to be fulfilled in order to ensure thermodynamic compatibility. In otehr words, the evolution of the system is thermodynamically consistent when the chemical potential becomes negative in the presence of increasing binding proteins on the membrane. An example of a similar instance was encountered in \cite{Gao:2005b} in which an \textit{ad hoc} flow rule respecting such condition was introduced.

\subsection{Constitutive equations for the membrane}
\label{sec:3M}
In presence of the uniaxial kinematics discussed in Section \ref{sec:planarcase}, the problem is entirely governed by a one-dimensional energetics. 

In agreement with \eqref{position}, a complete representation formula for the Helmoltz free energy density introduced in \eqref{energy-reduced-reference} is sought. Indeed $W=W(\lambda,\lambda_x, n_i)$ represents the energy per unit area in the active (reference) configuration 
%can be 
%is considered
%, say $W=h_0W^*=W(\lambda,\lambda_x, n_i)$, 
for which no complete expression is yet available. With the aim of providing an explicit representation for %, the energy density 
$W$, 
%connected with the dynamically evolving lipid membrane, 
an additive decompositon is assumed  between the (hyperelastic)energy of the membrane and a second potential associated to the energetics of the transmembrane proteins.
%here decomposed in the difference between a first potential describing the effective hyperelastic response of the membrane and a second potential associated to the energetics of the transmembrane proteins, for which nonlocal effects are excluded. 
This assumption is stated in the following form:
\begin{equation}\label{W01D}
W=W_{hyp}(\lambda,\lambda_x)-W_{n_i}(\phi,n_i)=W_{hyp}(\lambda,\lambda_x)-W_{n_i}(\lambda^{-1},n_i).
\end{equation}
it is worth notning that the second term account for the possibility %where it is provided 
that activation and conformational changes of GPCRs may affect the membrane thickness, measured through $\lambda^{-1}$. The total differential of $W$ thus gives:
\begin{equation}\label{W01Dvar}
dW=\frac{\partial W}{\partial \lambda} d\lambda+\frac{\partial W}{\partial \lambda_x} d\lambda_x + \frac{\partial W}{\partial n_i} dn_i= \tilde{P}\,d\lambda + \tilde{\mu}_i\, dn_i.
\end{equation}
Because of relations \eqref{ConstEqual}, the following equations for the stress and for the chemical potentials relative to the active (reference) configurations follow:
\begin{align}
&\tilde{P}=K_r^{-1}P=\frac{\partial W}{\partial \lambda}-\frac{\partial}{\partial x}\frac{\partial W}{\partial \lambda_x}\label{costPmu1}\\
&\tilde{\mu}_i=K_r^{-1}\mu_i= \frac{\partial W}{\partial n_i}=-\frac{\partial W_{n_i}}{\partial n_i}.\label{costPmu2}
\end{align}
A Legendre transformation of \eqref{W01Dvar} can now be performed by subtracting the variation $d(n_i\tilde{\mu}_i)=\tilde{\mu}_i\, dn_i+ n_i\,d\tilde{\mu}_i$, by obtaining the coupled potential \color{black}
\begin{equation}\label{W01Dvar2}
d\hat{W}=d(W_{hyp}(\lambda,\lambda_x)-\hat{W}_{\mu_i}(\lambda^{-1},\mu_i))= \tilde{P}\,d\lambda -  n_i\,d\tilde{\mu}_i
\end{equation}
from which it follows that 
\begin{equation}\label{niW}
n_i=-\frac{\partial \hat{W}}{\partial \tilde{\mu}_i}=\frac{\partial \hat{W}_{\mu_i}}{\partial \tilde{\mu}_i}.
\end{equation} 
Because $W_{hyp}$ is independent of the chemical potentials \color{black} and since $\hat{W}_{\mu_i}$ depends on the variables $\lambda^{-1}$ and $\tilde{\mu}_i$, the latter relation also implies that \color{black}
\begin{align}\label{dniW}
dn_i&=-d\left(\frac{\partial \hat{W}}{\partial \tilde{\mu}_i}\right)=\frac{1}{\lambda^{-2}}\frac{\partial\tilde{P}}{\partial \tilde{\mu}_i}\,d\lambda^{-1} - \frac{\partial^2\hat{W}}{\partial \tilde{\mu}_i^2}\,d\tilde{\mu}_i= - c_{\lambda_i}\,d\lambda^{-1} + c_{\mu_i}d\tilde{\mu}_i.
%-\frac{\partial}{\partial \tilde{\mu}_i}\left[d(W_{hyp}(\lambda,\lambda_x)-\hat{W}_{\mu_i}(\lambda^{-1},\tilde{\mu}_i)) \right]=%%\notag\\
%&=\frac{\partial^2\hat{W}_{\mu_i}(\lambda^{-1},\tilde{\mu}_i) }{\partial \lambda^{-1}\partial\tilde{\mu}_i}\,d\lambda^{-1}+ %%%%\frac{\partial^2\hat{W}_{\mu_i}(\lambda^{-1},\tilde{\mu}_i) }{\partial\tilde{\mu}_i^2}\,d\tilde{\mu}_i. \notag\\
%&= - c_{\lambda_i}\,d\lambda^{-1} + c_{\mu_i}d\tilde{\mu}_i.
%-\frac{\partial^2 W_{hyp}(\lambda)}{\partial \tilde{\mu}_i\partial\lambda}\,d\lambda + .
\end{align}
\color{black}
\noindent Herein, the coefficient $c_{\lambda_i}$ and $c_{\mu_i}$ relate the variation of the $i^{th}$ species density $n_i$ to the change of the membrane stretch and of the associated chemical potential $\tilde{\mu}_i$.
It is assumed that the terms $c_{\lambda_i}$ and $c_{\mu_i}$ can be modelled as first order variation coefficients that depend on the current amount of active proteins, i.e. $c_{\lambda_i}\approx n_i \overline{c}_{\lambda_i}$ and, in the light of the antagonism \eqref{localdissip2}, $c_{\mu_i}\approx -n_i \overline{c}_{\mu_i}$. Under this constitutive assumption, equation \eqref{dniW} rewrites as:
\begin{equation}\label{dniW2}
\frac{dn_i}{n_i}= - \overline{c}_{\lambda_i}\,d\lambda^{-1} - \overline{c}_{\mu_i}d\tilde{\mu}_i
\end{equation}
Therefore, \color{black}
% by considering that moderate variations of protein densities induce small variations of stretch, and consequently of membrane thickness, as well as of the related chemical potentials, 
integration of Equation \eqref{dniW2} leads to the following relation:\color{black}
%it is possible to introduce following approximate constitutive relation:
\begin{equation}\label{niRel}
-\log \frac{n_i}{n_i^0}\simeq \, \overline{c}_{\lambda_i}\,\left(\lambda^{-1}-1\right) +\overline{c}_{\mu_i}\left(\tilde{\mu}_i-\tilde{\mu}_i^0\right)
\end{equation}
which allows for evaluating the chemical potential of the species as follows
\begin{equation}\label{ChemPot}
\tilde{\mu}_i=\tilde{\mu}_i^0 - \omega_i \, \left(\lambda^{-1}-1\right) -\eta_i \, \log \frac{n_i}{n_i^0},
\end{equation}
\color{black} after setting 
\begin{align}
&\omega_i := \overline{c}_{\lambda_i}/\overline{c}_{\mu_i},\notag\\
\label{Chemical_Potential_Coefficients}\\
&\eta_i := 1/\overline{c}_{\mu_i}. \notag
\end{align}
%$\omega_i=\overline{c}_{\lambda_i}/\overline{c}_{\mu_i}$ and $\eta_i=1/\overline{c}_{\mu_i}$. 
%
Equation \eqref{costPmu2} in combination with \eqref{ChemPot} then allows for finding the following (normalized) expression for the potential $W_{n_i}$:
\begin{equation}\label{Wnidef}
W_{n_i}(\lambda^{-1},n_i)=-\int_{n_i^0}^{n_i}\tilde{\mu}_i\,dn_i= \eta_i \, n_i \, \log \frac{n_i}{n_i^0}+(n_i-n_i^0)\left(\omega_i\left(\lambda^{-1}-1\right) -\tilde{\mu}_{i}^0 -\eta_i \right).
\end{equation}
This relation exhibits a Boltzmann-type entropic contribution related to the chemical species coupled with a mechanical term. In \hyperref[{lateralpress}]{Appendix B} it has been shown that the latter represents the specific work done by the lateral pressure between the protein domains and the lipids in the cell membrane during conformational changes.

Representation formula \eqref{Wnidef} for $W_{hyp}$ permits to obtain the following relationships for the stress and the chemical potentials:
\begin{align}
&P=K_r\left(\frac{\partial W_{hyp}}{\partial \lambda}-\frac{\partial}{\partial x}\frac{\partial W_{hyp}}{\partial \lambda_x}-\frac{\partial W_{n_i}}{\partial \lambda}\right)=\notag\\
&\quad=K_r\left(\frac{\partial W_{hyp}}{\partial \lambda}-\frac{\partial}{\partial x}\frac{\partial W_{hyp}}{\partial \lambda_x}-\omega_i  \frac{(n_i^0-n_i)}{\lambda^2}\right)=P_{eff}-P_{n}, \label{Pmudef1}\\
&\mu_i = K_r \tilde{\mu}_i=K_r\left(\tilde{\mu}_i^0 - \omega_i \, \left(\lambda^{-1}-1\right) -\eta_i \, \log \frac{n_i}{n_i^0}\right). \label{Pmudef2}
\end{align}
\color{black}
In equation \eqref{Pmudef1}, $P$ can be interpreted as the membrane net stress, while $P_{eff}$ and $P_n$ respectively represent the effective (hyperelastic) response of membrane and the \textit{chemical stress} due to protein specific work. \color{black} Indeed, the assumed correlation between the potential associated to transmembrane proteins activation and lipid bilayer thickening has been introduced in equation \eqref{W01D}. It can be noted that each term $\omega_i (n_i^0-n_i)/ \lambda^2$ forming $P_n$ relates to the change of the corresponding specific work exerted by the lateral pressure arising in species-lipid interactions. In \hyperref[{lateralpress}]{Appendix C} this is explicitly worked out for the active receptors through a bottom-up approach that considers the surface interactions between receptors and lipids, thereby giving a mechobiological explanation to recently observed experimental findings.\color{black} The effective response of the lipid membrane is instead fully identifiable with the hyperelastic law \eqref{eqB:energy_gamma_1D} for $J=\lambda$, here reported for convenience:

\begin{equation}\label{DZpot}
W_{hyp}(\lambda) := \psi(\lambda)=\varphi(\lambda) -\frac{1}{2}\gamma(\lambda)(\lambda_x)^2
\end{equation}

\subsection{Mechanical equilibrium of the biological cell membrane}
\label{sectionD0}
In this Section, the use of total free-energy \eqref{W01D} expressed by means of  \eqref{Wnidef} and \eqref{DZpot} as a simplified energy governing the behaviour of biological membranes allows for searching the set of mechanical equilibria. To this end, the variational derivative of the resulting Gibbs free energy with respect to the in-plane displacement $u$ will be computed in order to isolate configurations corresponding to its stationary points. The Gibbs free energy \color{black}, here denoted by $\mathcal{E}$ \color{black} is obtained as \color{black} the difference between the Helmholtz one $\mathcal{H}$, defined by \eqref{eq:energy_basic}$_2$ below,\color{black} minus the work $\mathcal{L}$ done by the external load, i.e
\begin{equation}
\label{eq:energy_basic}
\mathcal{E} :=\mathcal{H} - \mathcal{L}, \qquad \mathcal{H}:=\int_\Omega W_0 d\Omega
\end{equation}
The work $\mathcal{L}$ results from the presence of two possible mechanical agents, a traction $\Sigma(t)$  acting against the displacement, and a hyperstress $H(t)$ acting against the gradient of the displacement at the boundary, i.e.
\begin{equation}
\mathcal{L} = \left[\Sigma\,u + H\, u_x\right]_{\partial\Omega},
\end{equation}
\color{black} Because $u(x,t)=g(x,t)-x$, since \eqref{eqB:lambda} holds, a compatibility relation between the stretch and the derivative of the displacement is obtained $\lambda= 1 + u_x$. Henceforth, upon considering \color{black} a perturbation $\delta u$, %of the displacement, 
the first variation of \eqref{eq:energy_basic} with respect to the displacement can be written % also 
 with the help of relation \eqref{dissip3}: 

\begin{align}\label{first_variation}
\delta\mathcal{E}&= \left[\frac{\partial W_0}{\partial \lambda_x}\,\delta u_x\right]_{\partial \Omega} +\int_\Omega P\,\delta u_x\, d\Omega - \left[\Sigma\,u + H\, u_x\right]_{\partial\Omega}\notag\\
&= \left[\left(\frac{\partial W_0}{\partial \lambda_x}-H\right)\,\delta u_x\right]_{\partial \Omega}+\left[\left(P-\Sigma\right)\delta u\right] -\int_\Omega P_x \,\delta u\, d\Omega.
\end{align}

Stationary condition of the energy \eqref{first_variation} supplies the following equations governing the equilibrium of the membrane: 
\begin{equation}\label{equilibrium}
\begin{cases}
 P_x =0, \quad x\in\Omega\\
\text{either}\,\, P=\Sigma\,\,\text{or}\,\, \delta u=0, \quad x\in \partial \Omega\\
\text{either}\,\, \gamma\,u_{xx}=H\,\,\text{or}\,\, \delta u_x=0, \quad x\in \partial \Omega
\end{cases}
\end{equation}
in which equation \eqref{DZpot} and compatibility relation have been employed, while \color{black} the stress $P$ relative to the virgin configuration reads as follows: 
\begin{equation}\label{StressDisp}
P=K_r\left[\tau(\lambda) + \frac{1}{2} \gamma'(\lambda) (u_{xx})^2+\gamma(\lambda) u_{xxx} - \omega_\xi \frac{\xi-\xi^0}{(1+u_x)^2} - \omega_\zeta \frac{\zeta-\zeta^0}{(1+u_x)^2}\right].
\end{equation}
Here \color{black} $\tau$ and $\gamma$ have been defined through \eqref{eq:tau_def}  and\eqref{eqB:gamma} respectively.
\subsection{Coupling of the equations}
For the sake of illustration, prediction the formation of lipid rafts on a cell membrane is obtained in the case of plane strain. Coupling between  the balance of linear momentum \eqref{equilibrium}$_1$ with the evolution laws provided by the mass balance of transmembrane proteins, \eqref{massbalance2.1} and \eqref{massbalance2.2}, yields the following set of governing equations:
\begin{equation}\label{Alleqs}
\begin{cases}
P=\Sigma\\
\dot{\xi}-\left(L_\xi\,\xi\,\mu_{\xi,x}\right)_x = \alpha_\xi \, \xi -\delta_\xi \, \xi -\beta_{\xi\zeta} \zeta \, \xi \\
\dot{\zeta}-\left(L_\zeta\,\zeta\,\mu_{\zeta,x}\right)_x= \beta_{\zeta\xi} \xi \, \zeta -\delta_\zeta \, \zeta .
\end{cases}
\end{equation}

Here, because of symmetry assumptions, $x\in\left[0, L\right]$ with $L$ set to $L=1000 h_0$ , while the time $t\in\left[0^+,t_{max}\right]$, $t_{max}$ being set to 1 hour on the basis the experiments duration (see Sect. appendix \ref{ExpMeas}). The unknowns of the problem are the displacement $u(x,t)$ and the protein fractions $\xi(x,t)$ and $\zeta(x,t)$, subjected to the boundary conditions $u(0,t)=0$, $u_{xx}(0,t)=u_{xx}(L,t)=0$, $\xi_x(0,t)=\xi_x(L,t)=0$ and $\zeta_x(0,t)=\zeta_x(L,t)=0$ in absence of hyperstress $H$ and protein species leakage at the boundaries. Initial conditions provide initial null displacement $u(x,0)=0$, while suitable initial distribution have been assigned to proteins. In particular, $\zeta(x,0)=\zeta^0$ is assumed constant, while, in order to take into account localization of lipid rafts on the membrane in response to prescribed ligand distributions, an initially non-homogeneous spatial profile of inactive GPCRs  has been introduced. To do this, the initial condition $\xi(x,0)=\xi^0(x)$ is expressed by means of a suitable generating function:
\begin{equation}\label{Xi0}
\xi^0(x)=\xi_o+ \left(\xi_a-\xi_o\right) \sum_{m=0}^{{m}_{raft}}\Pi(x), \qquad \Pi(x)=\frac{E_{-}(x)}{1+E_{-}(x)}-\frac{E_{+}(x)}{1+E_{+}(x)}
\end{equation}

where $E_{\pm}(x)=\exp\left[-c_a\left(x-x_a \pm \Delta_a/2\right)\right]$, $x_a$ and $\Delta_a$ governing the positioning and the amplitude of the activating raft. In particular, the formation of a single raft region centred at the origin with amplitude $\Delta_a=400h_0$ and multiple (quasi-periodic, with ${m}_{raft}=4$) rafts regions with amplitude $\Delta_a=100h_0$ have been simulated in the sequel. Moreover, a stress-free (i.e. $\Sigma=0$) and a stress-prescribed case have been both considered, by imposing that the nominal traction equals the Maxwell stress of the membrane through the expression  $\Sigma(t)=\Sigma_M(1-e^{-t})$, consistently with the initial undeformed conditions and with characteristic values of internal cell pressures \cite{cellpress}. Equations \eqref{Alleqs} have been numerically solved with the aid of the software Mathematica\textsuperscript{\textregistered} \cite{MathematicaProgram} through the method of lines. Spatial coordinates and model parameters, reported in \hyperref[{tab}]{Table} \ref{tab},  have been normalized with respect to the reference thickness $h_0$.

\begin{table}[]\label{tab}
\centering
\caption{\textbf{Table} \ref{tab}. Coefficients used in numerical simulations (e.d.= available experimental data).}
\label{tab}
\begin{tabular}{cccc}
\hline
Coefficient        & Value {[}Unit{]}       & Range {[}Unit{]}        & Reference \\ 
\hline
$k_b$              & $5.18$                  &$3.89-5.7$   &  \cite{Bridge2018d1, Li2017} - e.d.    \\
$\delta_\xi$       & $1.1\times10^{-3} \, s^{-1}$ &$(0.9-1.65)\times10^{-3} \, s^{-1}$ & \cite{Bridge2018d1}     \\
$\beta_{\xi\zeta}$ & $4\times10^{-6}\, s^{-1}$  &$(3-6)\times10^{-6} \, s^{-1}$  & \cite{Bridge2018d1, Rich2001}     \\
$\beta_{\zeta\xi}$ & $3.3\times 10^{-3}\, s^{-1}$ &$(2.5-4.15)\times10^{-3} \, s^{-1}$ & \cite{Saucerman2013, Rich2001, Agarwal2014} - e.d.   \\
$\delta_\zeta$     & $10^{-7}$            &$10^{-8}-10^{-6}$        &           assumed \\
$\zeta^0$          & $10^{-2}$            &$ $         &         -   \\
$\xi_o$            & $10^{-4}$              &$ $      &          - \\                     
$\xi_a$            & $10^{-2}$              &$ $      &          - \\ 
$L_i$            & $1.87\times10^{-17}\,{m^4 J^{-1}s^{-1}}$ &$(10^{-20}-10^{-15})\,{m^4 J^{-1}s^{-1}}$    &      \cite{gurevich2018gpcrs, Kim2016}  \\ 
\hline
\end{tabular}
\end{table}

\section{Results and discussion}
The outcomes of the numerical simulations derived by the biomechanical model introduced above show the lipid bilayer undergoing dynamic remodelling triggered by the evolution of transmembrane protein activity. In particular, here strain localization within the cell membrane leading to the coexistence of thicker (ordered $L_o$-phases) and thinner lipid zones (disordered $L_d$-phases) is predicted. 

\begin{figure}[htbp]
\centering
\includegraphics[width=0.99\columnwidth]{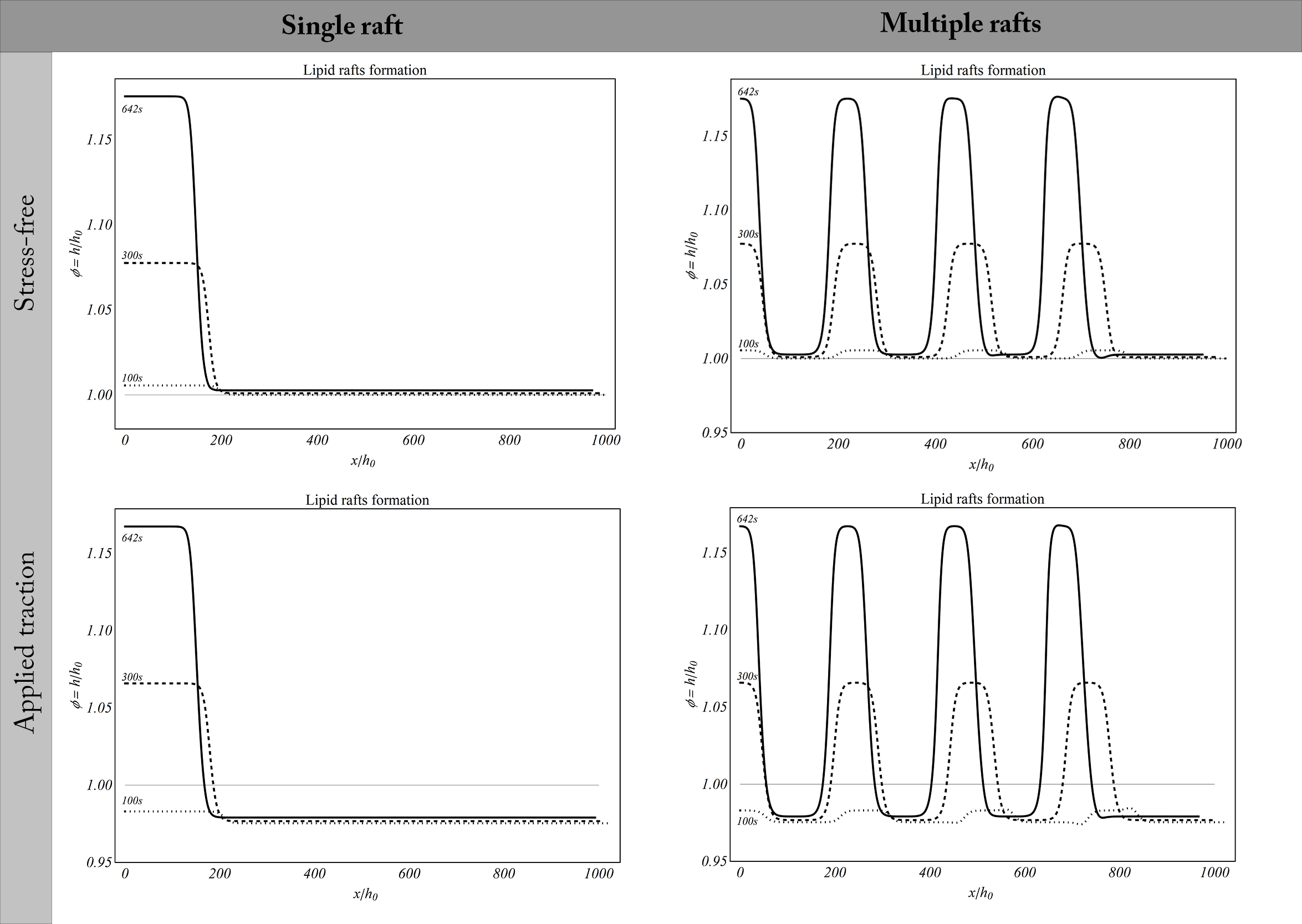} 
\caption{Lipid raft formation in stress-free (upper row) and tensed (lower row) membranes. Both the case of a single raft and multiple raft are analysed, the membrane thickening following the activation of transmembrane proteins.}
\label{fig:Phi}
\end{figure}

As shown in Figure 
\ref{fig:Phi}, \color{black} the densification of lipids across the regions in which active GPCRs tend to cluster %during the simulations 
is accompanied by the progressive formation of lipid rafts. In agreement with previous findings (see e.g. \cite{niemela2007}), on such sites a difference of about $0.9$ nm between raft and non-raft membrane zones is detected. \color{black} In particular, in stress-free simulations, there is a pure chemically-induced thickening of the membrane that forms a single raft and multiple rafts according to the analysed cases (Figure \ref{fig:Phi}). Under the same initial conditions, rafts also occur in the tensed membrane, where a transverse contraction and initial longitudinal elongation additionally take place in the $L_d$-phase as elastic response of the applied traction. 

\begin{figure}[htbp]
\centering 
\includegraphics[width=0.5\columnwidth]{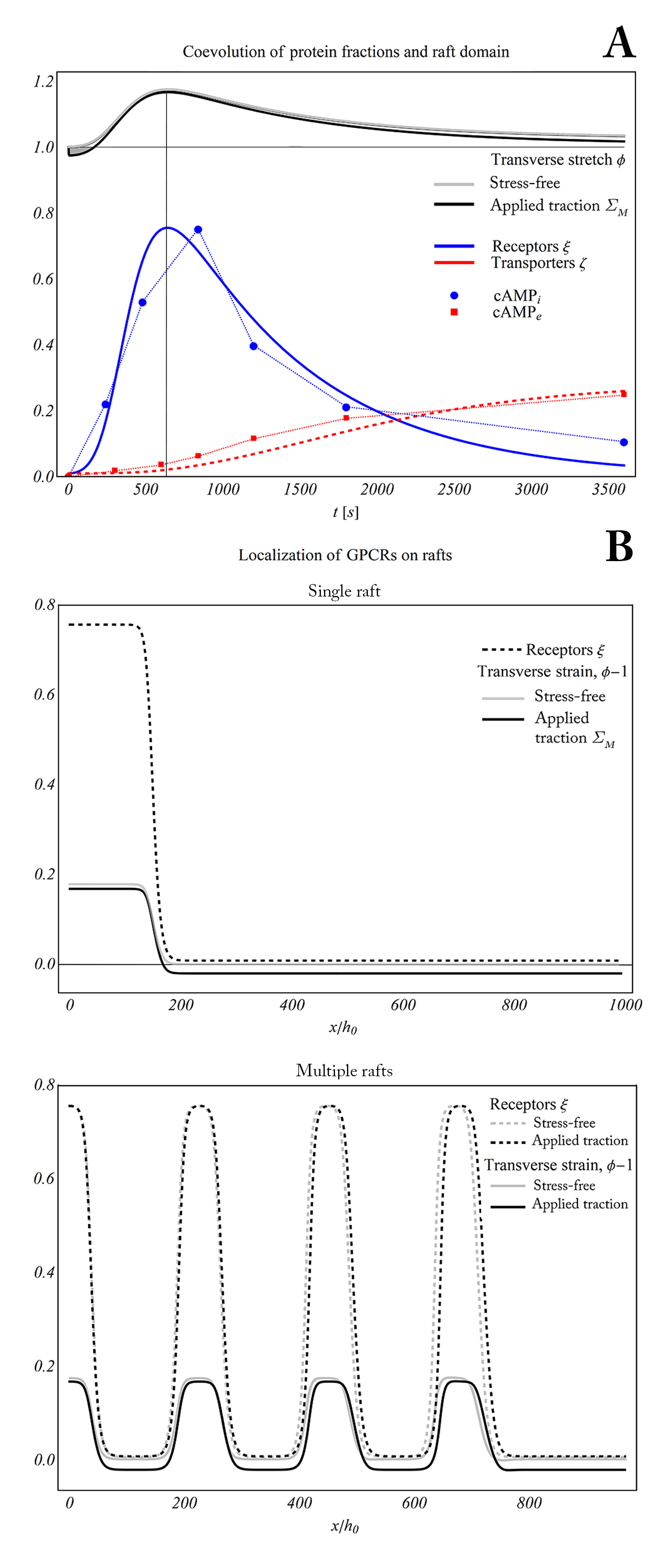} 
\caption{\textbf{A}. Time evolution of active receptors (blue) and transporters (red) in comparison with the experimental cAMP measurements. It is shown that the raft dynamically follows the binding/unbinding kinetics. \textbf{B}. High degree of localization between lipid raft formation and active GPCRs in single rafts (top) and multiple raft domains (bottom).}
\label{fig:XiPhi}
\end{figure}

In Figure \ref{fig:XiPhi}\hyperref[{fig:XiPhi}]{A} the time-evolution of both the type of transmembrane protein fractions \textcolor{black}{occurring on a raft site} is reported. The interplay between RL proteins $\xi$ and transporters $\zeta$ is studied as a result of the introduced interspecific dynamics. More specifically, an initial growth of the GPCRs binding with ligands first occurs. Upon comparing this finding with the experimentally determined intracellular cAMP (cAMP\textsubscript{i}) normalized concentration, an estimate of the binding Receptor-Ligand (RL) compounds is provided according to \eqref{eq:camptoxi}. This is because GPCRs activation directly triggers the cell response by initiating the cAMP signaling pathway. Homeostatic levels of cAMP\textsubscript{i} are restored thanks to the activation of the Multidrug Resistant transport Proteins (MRPs) driving the efflux of part of the produced cAMP towards the extracellular environment. %, this establishing a basically proportional dependence. 
It is worth noting that a very good agreement is found between the predicted MRPs fraction and the (normalized) increase of cAMP\textsubscript{e}. 

Furthermore, Figure \ref{fig:XiPhi}\hyperref[{fig:XiPhi}]{A} displays the simultaneous evolution of the ordered (thicker) phases. The latter evolve with similar trends in both the unconfined and the tensed membranes, exception made for the load-induced shrinkage in the initial stage, when chemical stress $P_{n} \, =\, K_r \,\omega_i  \, (n_i^0-n_i)/\lambda^2$ (introduced in \eqref{Pmudef1}) is not yet acting. At the time at which such stress is maximum, i.e $t\simeq 642 s$ in the simulations, a clear spatial correspondence between active domains and raft formation can be recognized in all situations, see Figure \ref{fig:XiPhi}\hyperref[{fig:XiPhi}]{B}. \textit{This predicts that active GPCRs localize on lipid rafts} as an effect of the chemo-mechanical coupling between: 

\begin{itemize}
\item the chemical pressure arising from the deformation of the lipid bilayer;\\
\item the configurational changes of TMs domains, to which a decrease chemical potential as in Figure \ref{fig:Mu} is associated according to the above discussed thermodynamic arguments.
\end{itemize}

\begin{figure}[htbp]
\centering
\includegraphics[width=0.99\columnwidth]{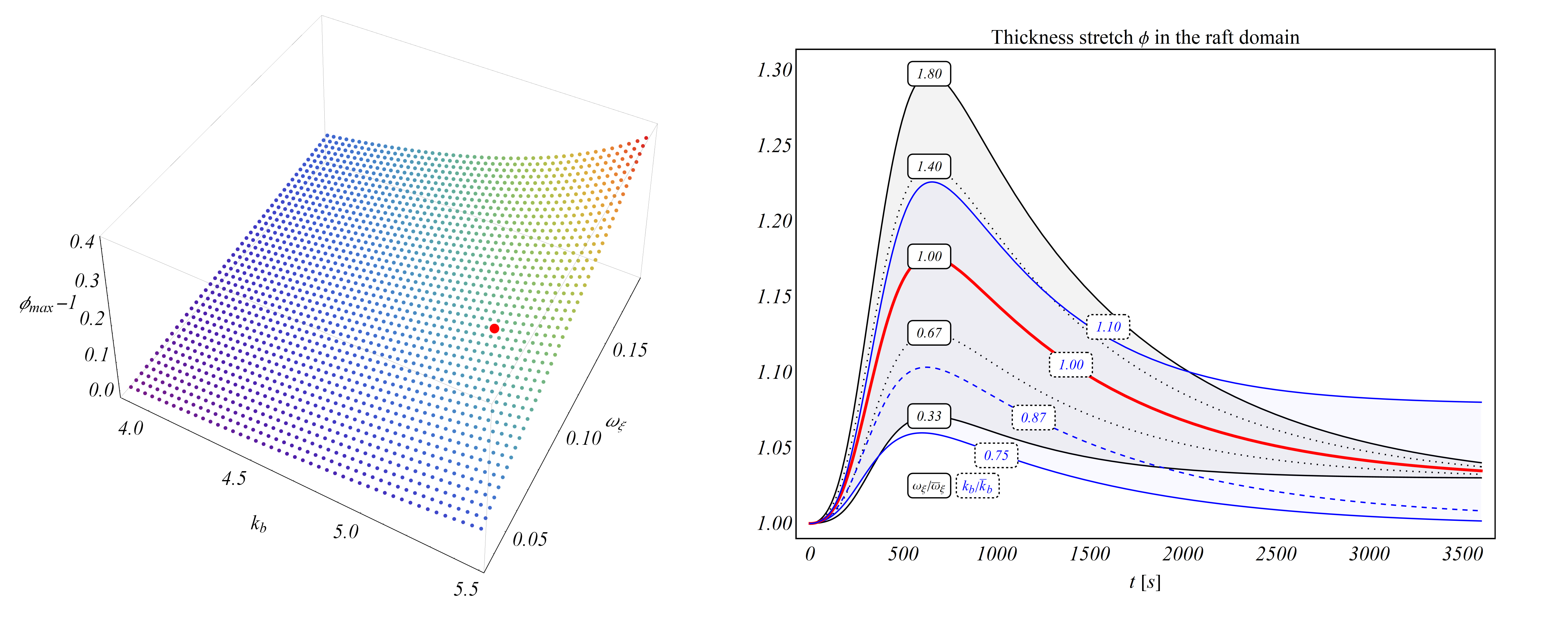} 
\caption{(Left) Variation of the maximum thickening of the lipid membrane as a function of the intrinsic uptake coefficient $k_b$ and of the chemo-mechanical work coefficient $\omega_\xi$. The red spot indicates the couple $\{\overline{ k }_b , \overline{\omega}_\xi \}$ used in the simulation. (Right) Evolution of the membrane stretch for different values of coefficients $k_b$ and $\omega_\xi$. The red curve highlights the one obtained for the adopted parameters.}
\label{fig:sensan}
\end{figure}

\begin{figure}[htbp]
\centering
\includegraphics[width=0.5\columnwidth]{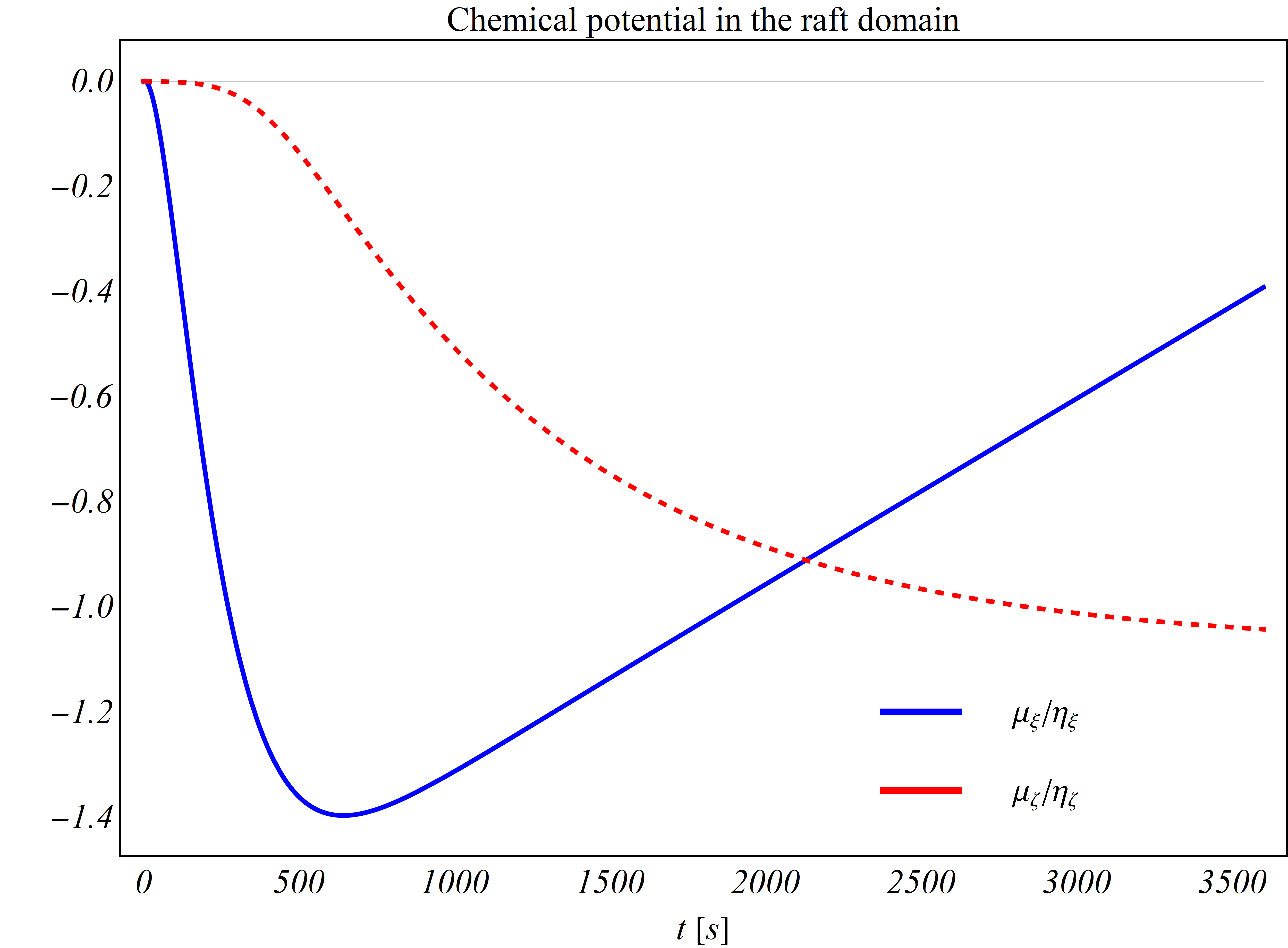} 
\caption{Evolution of chemical potentials associated to protein species during binding/unbinding kinetics.}
\label{fig:Mu}
\end{figure}

Although the model and the associated numerics is carried out for a one-dimensional problem, the findings are fully consistent with the experimentally observed clustering of ligand-binding receptors on lipid rafts across the cell membrane (see e.g. \cite{Bray:1998, foster2003, Chini:2004, Ostrom:2004, Zhang:2005, Becher:2005, Watkins:2011, Villar:2016} among others). 
\textcolor{black}{In order to investigate how model coefficients influence the formation of lipid rafts in high coupling with the dynamics of active receptors, the robustness of the model was tested by considering the variation of both interspecific and chemo-mechanical coefficients. In this regard, more than two thousands numerical simulations were performed by assigning values within the ranges reported in Table \ref{tab} and by then evaluating the associated membrane thickening. In particular, as also shown in Figure \ref{fig:sensan}, analyses revealed that lipid rafts formation is mainly affected by the level of intrinsic rate of activation in response to ligand concentration $\alpha_\xi$ (through the uptake coefficient $k_b$), and the chemo-mechanical coefficient $\omega_\xi$, representing the specific mechanical work that GPCRs exert on the lipid membrane by inducing its thickening. }

\begin{figure}[htbp]
\centering
\includegraphics[width=0.99\columnwidth]{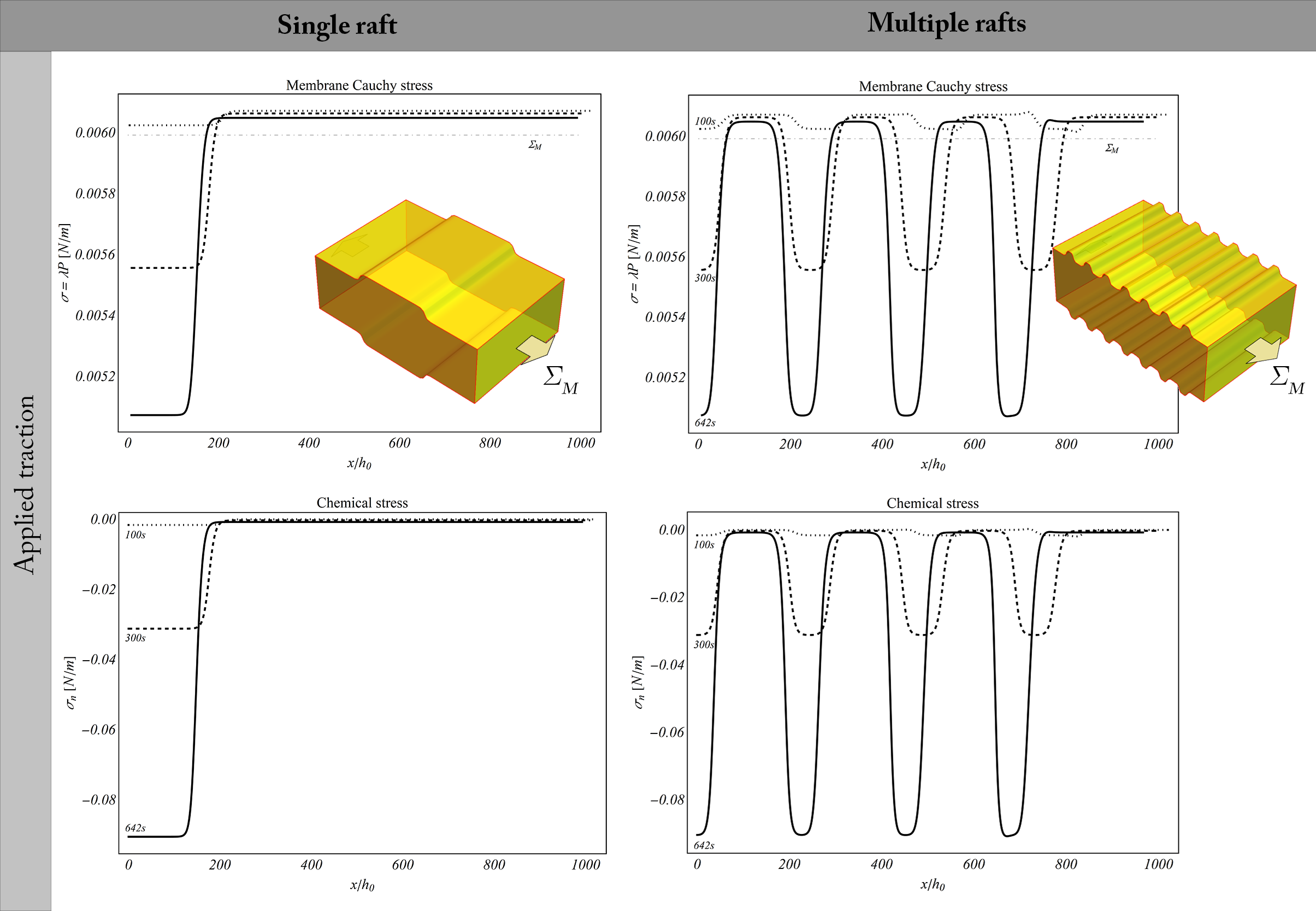} 
\caption{Stress in tensed membranes. The imposed constant reference traction induces a nonhomogeneous distribution of the Cauchy stress (upper row), which relaxes in thickened raft regions. In the lower row, the chemical stress generated by the protein species is reported, causing a progressive compression of the lipids inhabiting the raft domains.}
\label{fig:StressSM}
\end{figure}

As displayed in Figure \ref{fig:StressSM}, the imposed constant reference traction induces a nonhomogeneous distribution of the Cauchy stress, which relaxes within the rafts (namely across the thickened regions), thereby exhibiting values lower than the tension experienced by the thinner (hence disordered) phases. The binding kinetics and protein activation are instead responsible of a progressive, chemically-induced, compression of the lipids in the raft region. \color{black} Such compression decreases over time during the unbinding phase (see Figure \ref{fig:StressSM}).  \color{black}

\begin{figure}[htbp]
\centering
\includegraphics[width=0.99\columnwidth]{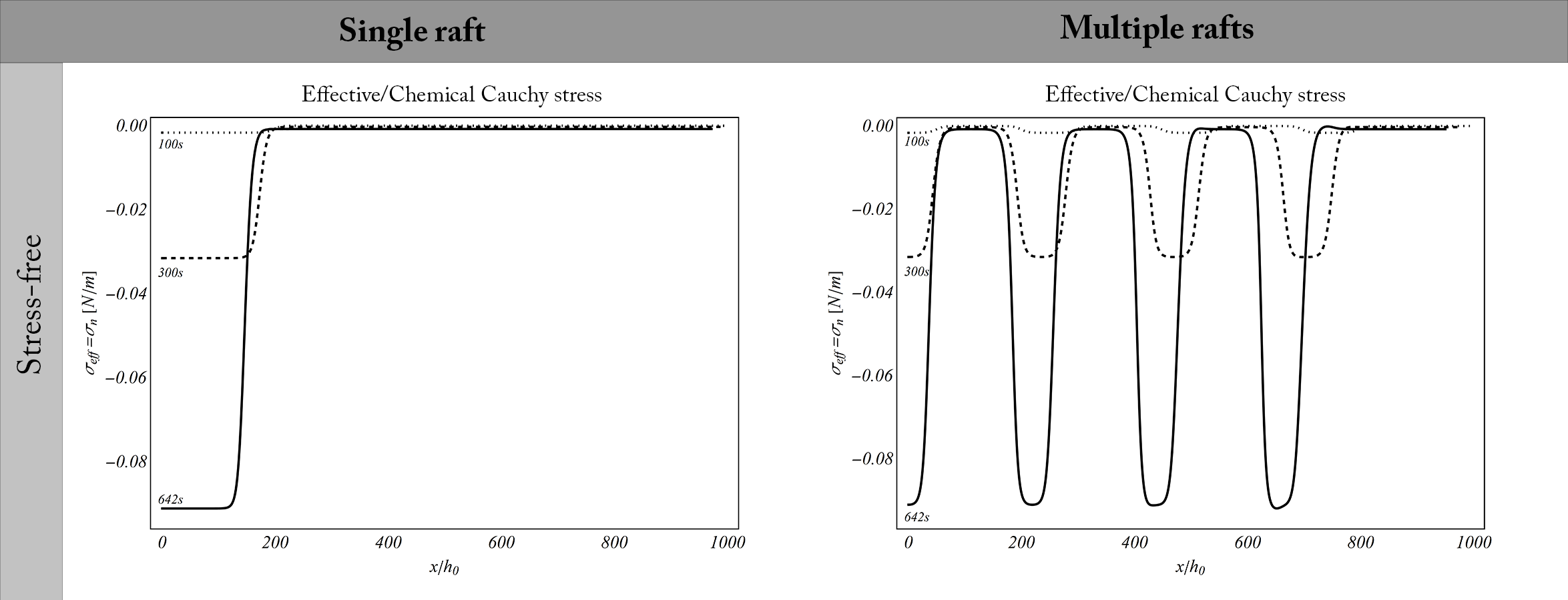} 
\caption{Effective and chemical stress balancing in stress-free remodelling membranes.}
\label{fig:Stress0}
\end{figure}

In stress free conditions, the chemical stress exerted by the activating proteins is entirely converted into effective membrane compression (see Figure \ref{fig:Stress0}). From a mechanical point of view, the stress relaxation renders ordered islands more compliant to demand TMs to undergo conformational changes, and hence GPCRs location on lipid rafts also implies a reduced energy expenditure during the protein structural re-organization.

% == FIGURE ==
% ======

% == FIGURE ==
% ======

% == FIGURE SENS AN ==
% ======

% == FIGURE ==
% ======

% == FIGURE ==
% ======

% == FIGURE ==
% ======

%==============================================
% Conclusion
%==============================================
\section*{Conclusions}
\label{sectionH0}
\color{black} Predictions and interpretation of the findings provided in this paper allow for explaining \textit{why active receptors} of the GPCRs family \textit{tend to cluster on lipid rafts}. Furthermore, \textit{active receptors are predicted to enhance rafts formation}. 

Such predictions are shown to be in relation with: \color{black}

i) stress reduction arising within rafts because of the presence of  active receptors and transporters that need local compliant zones to be able to perform conformational changes;

ii) decrease of chemical potential owing (1) binding of the GPRCs with the ligand, thereby favoring the formation of higher density regions of RL complex and (2) uptaking of intracellular cAMP from the MRPs transporters.

%The interplay among energetics, kinetics of binding and unbinding of GPCRs and MRPs is obtained. 
It is shown how the mechanobiology involved at the level of the cell membrane does entail strong coupling among (a) mechanical equilibrium governed by the (entropic, conformational, elastic) energy, (b) the kinetics of RL-binding and unbinding, regulated by the chemical potential and traced through the diffusion \color{black} of  GPCRs and MRPs (related to cAMP$_i$ and cAMP$_e$ respectively), \color{black} and (c) the consistency requirement that the effective chemical potential of the RL-compound must agree both with the energetics and with the fact that lateral pressure arising between lipids surrounding \color{black} the domains of the active species \color{black} and confining their conformational changes. The strict correlation between how cAMP is triggered by RL-compounds \color{black} and modulated by MRPs and the tendency of GPCRs to locate on lipid rafts has been shown by the simulations. %The total energy has been derived by coupling the energy of the hyperelastic membrane and the energetics for the species involved in the process, namely the active GPCRs and MRPs. This allows to find both a suitable expression for the chemical potential and the chemical stress, both associated to the work done by the lateral pressure exerted by the species domains against the lipid membrane. 
\textcolor{black}{It is worth notning that on the basis of the boundary and limit conditions of interest and by considering the interspecific terms taken into account in the model, coalescence due to Cahn-Hilliard-like dynamics was not expected, this phenomenon being object of future investigations.} \color{black}
The entire analysis has be done with special regard to an existing set of experiments involving HTR-8/SVneo cells and $\beta$2AR. Diffusive phenomena have been considered and the arising chemo-mechanical coupling has been accounted for through the kinetics of ligand binding. The outcomes of the model highlight the mechanobiology of lipid membrane remodelling by providing a mechanical explanation for the observed correspondence between the membrane regions where distribution density of active receptors is higher and locations of lipid rafts. These theoretical predictions actually confirm experimental findings by Nobel Prize 2012 Kobilka \cite{Kobilka:2007} and Lefkovitz \cite{Lefkowitz:2007}. Importantly, the predictive potential of the obtained model can help to provide a key support to quantitative diagnostics, owing for evaluation of cell response to endogenous and exogenous ligands regulated by GPCRs under different operating conditions of the cells. 

%\section*{Acknowledgements}
%L.D. gratefully acknowledges the partial support from the grant (1) ERC-2013 - ADG-340561- INSTABILITIES. A.C.,N.P., L.D. and M.F. acknowledge the partial support from the grants (2) ARS01-01384-PROSCAN and (3) PRIN 2017 20177TTP3S. N.P. and L.D. also ackowledge the partial support from (4) the Italian MIUR "Department of excellence grant" L. 232/2016 and (5) H2020 FET Proactive project NEUROFIBRES.  K. D. acknowledges support from (6, 7) NSF Mechanics of Materials and Structures (1150002 and 1635407), (8) NSF Manufacturing Machines andEquipment (1635435), (9) ARO Numerical Analysis (W911NF-17-1-0084), and (10) ONR Applied and Computational Analysis (N00014-18-1-2528).

\section*{Acknowledgements}

L.D., N.M.P. and M.F. acknowledge the Italian Ministry of Education, University and Research (MIUR) under the (1) ARS01-01384-PROSCAN and (2) PRIN 2017 20177TTP3S grants. 
N.M.P. and L.D. also acknowledge the partial support from (3) the Italian MIUR "Departments of Excellence" grant L. 232/2016. 
N.M.P. and L.D. are supported by the European Commission (EC) under (4) the FET Proactive (Neurofibres) grant No. 732344. 
%N.M.P. also acknowledges support from the EC under the (5) Graphene Flagship Core 2 grant No. 785219 (WP14, Composites) and (6) the FET Open (Boheme) grant No. 863179. \color{black}
A.R.C. acknowledges support from (5) PON-AIM1849854-1. 
M. B. and K. D. acknowledge support from (6, 7, 8) NSF (1150002, 1635407 and 1635435), (9) ARO (W911NF-17-1-0084), (10) ONR Applied and Computational Analysis (N00014-18-1-2528), and (11) BSF (2018183).  
The authors wish to acknowledge Prof. Carla Biondi, Prof. Maria Enrica Ferretti and Prof. Fortunato Vesce, from the University of Ferrara, Italy,  for the endless conversations about the role of cAMP, its detection and its relationships with submacroscopic effects at the level of bound receptors, and their help to Dr. Laura Lunghi during her experimental work. Prof. Giuseppe Valacchi, from North Carolina State University, USA, and Ferrara, is also acknowledged.
The authors also gratefully acknowledge Prof. Davide Bigoni and Dr. Diego Misseroni, from the University of Trento (Italy), for the discussions about biomechanics of cancer in relation to the findings of this paper, as well as Prof. Anne M. Robertson, from the University of Pittsburgh), Prof. David R. Owen, from Carnegie Mellon University, and Dr. Giuseppe Zurlo, from NUI-Galway-Ireland, for their valuable discussions through the last few years about various biological and theoretical aspects of this paper.

\bigskip

%
% FORMERLY APPENDIX A WAS HERE-EXPERIMENTAL SET UP
%
%%%%%%%%%%%%%%%%%%%%%%%%%%%%%%%%%%
\section*{Appendix A} \label{KrCalc}

\subsection*{Density ratio associated to remodelling}
\renewcommand{\theequation}{A.\arabic{equation}}
  % redefine the command that creates the equation no.
\setcounter{equation}{0}  % reset counter
%%%%%%%%%%%%%%%%%%%%%%%%%%%%%%%%%%%%%%%%%%%
The \color{black} effective density $\rho(x,t)$ of the heterogeneous membrane at any time $t$ and at any point $x$ \color{black} can be written as the sum of the products between the true densities $\varrho_k$ and the volumetric fractions $\upsilon_k$ composing the body. \color{black} The membrane is %basically 
essentially inhabited by lipids and by transmembrane proteins. The latter share the same true density, although they can appear either in an active or inactive state. Furthermore, it is worth notning that in the intermediate active (reference) configuration displayed in figure \ref{fig:Stress00}) the thickness of the membrane $h_0$ is homogeneous. Henceforth, the following expression can been written for the density $\rho(x,t)$: \color{black}
\begin{align}\label{dens1}
\rho(x,t)&= \upsilon_k\,\varrho_k=\upsilon_{u}\,\varrho^p_{u}+\upsilon_{a}\,\varrho^p_{a}+\upsilon_{l}\,\varrho^{l}\notag\\
&=\left(N_{u}\,\frac{A_{u}}{A}+N_{a}\,\frac{A_{a}}{A}\right)\varrho^p+\left[1-\left(N_{u}\,\frac{A_{u}}{A}+N_{a}\,\frac{A_{a}}{A}\right)\right]\varrho^{l}\notag\\
&=\varrho^l\left[1+\left(N_{u}\,\frac{A_{u}}{A}+N_{a}\,\frac{A_{a}}{A}\right)\left(\frac{\varrho^p}{\varrho^l}-1\right)\right]
\end{align}

where $\varrho^p=dm^p/dV^p$ and $\varrho^{l}=dm^l/dV^l$ are the protein and lipid true densities, respectively, while $A_k/A$ is the areal fraction of the species. By introducing the relation $N_u+N_a=N_{max}$, \color{black} where $N_{max}$ denotes an estimate of the  maximum reference value of total proteins present on a surface surface $A$ on the cell membrane, the expression \eqref{dens1} above can be rearranged as follows: \color{black}
\begin{align}\label{dens2}
\rho(x,t)&=\varrho^l\left[1+N_{max}\left(\frac{A_{u}}{A}+\frac{N_a}{N_{max}}\left(\frac{A_a-A_u}{A}\right)\right)\left(\frac{\varrho^p}{\varrho^l}-1\right)\right]\notag\\
&=\varrho^l\left[1+\frac{N_{max}A_u}{A}\left(1+\sum_i\,n_i\left(\frac{A_a}{A_u}-1\right)\right)\left(\frac{\varrho^p}{\varrho^l}-1\right)\right]
\end{align}
where $\sum_i\,n_i=N_a/N_{max}=\xi+\zeta$ represents the cumulative fraction of activated \color{black} receptor and transporters\color{black}. Therefore, the density ratio associated to remodelling, reported in equation \eqref{RemodelingTerm}, is finally obtained as  $K_r={\rho(x,0)}/{\rho(x,t)}$. \color{black} It is worth notning that $K_r$ has a meaningful multiscale geometric interpretation through the \textit{Theory of Structured Deformations}, as highlighted in Fig.\ref{fig:Stress0}  (see e.g. \cite{deseri:2003}, \cite{Deseri:2010}, \cite{deseri:2012}, \cite{deseri:2015}, \cite{deseri:2019}, \cite{palumbo:2018}), as discussed in details in \hyperref[{SB2}]{Appendix C}.\color{black}
%\color{black} As noted before, a revealing gometric interpretation for such a field is available by emplyoing the \textit{Theory of Structured Deformations}, as indicated in Fig.\ref{fig:Stress0} and analyzed in more details in \hyperref[{SB2}]{Appendix C}.\color{black}
%%%%%%%%%%%%%%%%%%%%%%%%%%%%%%%%%%%%%%%%%%%
\section*{Appendix B} \label{lateralpress}
\renewcommand{\theequation}{B.\arabic{equation}}
  % redefine the command that creates the equation no.
\setcounter{equation}{0}  % reset counter
%%%%%%%%%%%%%%%%%%%%%%%%%%%%%%%%%%%%%%%%%%%
\subsection*{The chemical potential is influenced by the lateral pressure experienced by the cell membrane}\label{lateral_pressure}
\renewcommand{\theequation}{B.\arabic{equation}}
  % redefine the command that creates the equation no.
\setcounter{equation}{0}  % reset counter
%%%%%%%%%%%%%%%%%%%%%%%%%%%%%%%%%%%%%%%%%%%
The mechanical term $\omega_i \, \left(\lambda^{-1}-1\right)$ appearing in the chemical potential $\tilde{\mu}_i=\tilde{\mu}_i^0 - \omega_i \, \left(\lambda^{-1}-1\right) -\eta_i \, \log \frac{n_i}{n_i^0}$, i.e. \eqref{ChemPot}, which is the parent of the chemo-mechanical energy \eqref{Wnidef}, allows for obtaining the chemical stress through derivation as in equations \eqref{Pmudef1} and \eqref{StressDisp}, i.e.
\begin{equation}\label{ChemStress}
\tilde{P}_n = (n_i^0-n_i) \frac{\partial \tilde{\mu}_i} {\partial \lambda} = - \frac{\partial W_{n_{i}}}{\partial \lambda} = \omega_i  \frac{(n_i^0-n_i)}{\lambda^2}.
\end{equation}
As the first formula clearly displays, \color{black} the chemical stress is obtained through the change in chemical potential caused by local stretch changes, \color{black} $\lambda$ (or, equivalently, caused by the local changes of the membrane thickness as \eqref{ThicknessChange}, i.e. $\phi=\lambda^{-1}$, so that $d\lambda=- \phi^{-2} d\phi$). The chemical potential of the species is then key for understanding the mechanobiology involved in ligand-binding of the receptors analyzed in this paper. Because of its high importance, in this section we provide a more physical understanding of the origins of such a key quantity. Indeed, direct considerations at the microscale can be performed by taking into account the lateral pressure arising between \color{black} the active species domains and the adjacent lipids within the bilayer. \color{black} Lateral pressure in lipid membranes is known to be present in bilayers even without receptors. It is worth noting that lateral pressure is known to contribute to the stress across the thickness. In mechanical terms, such a pressure can be seen as the reaction to incompressibility, thereby not affecting the elastic part of the energy. When it comes to considering the membrane stress, formed by a hyperelastic part added on a reactive one, the lateral pressure is indeed such a term in the expression for the stress.
In turn, the lateral pressure changes in the presence of proteins (see e.g. \cite{Cantor:1999, Lee:2004, Marsh:2007, Ollila:2010}) and it influences their chemical  potential. \color{black} 

\begin{figure*}[th]
\centering
\includegraphics[width=0.75\textwidth]{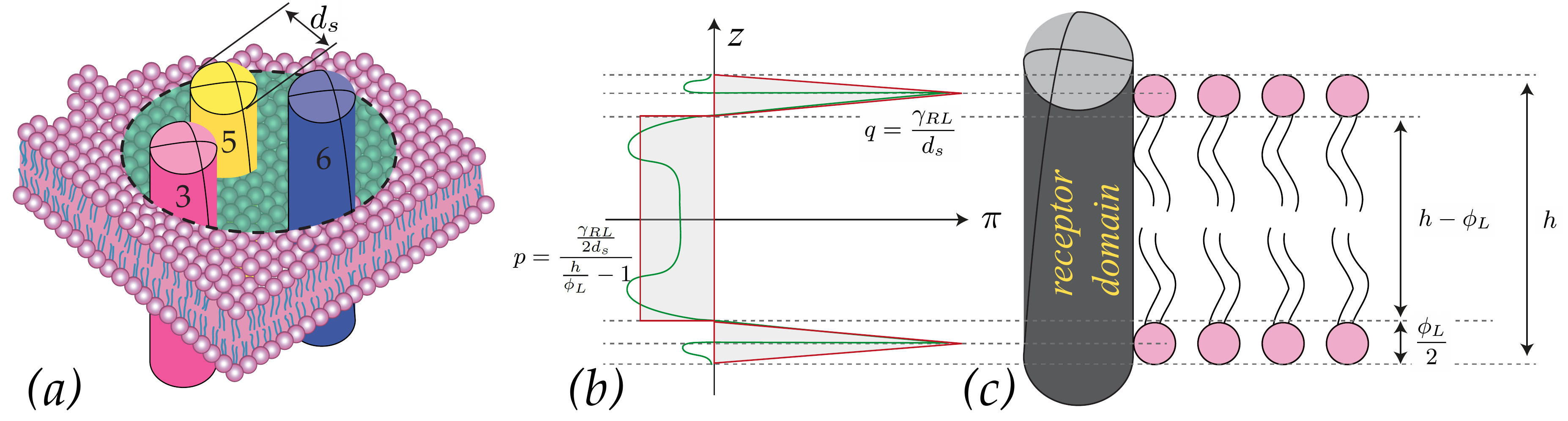}
\caption{Schematic representation of the lateral pressure between a TM domain and the lipid bilayer as in \cite{Pollaci:2016} (see also \cite{Marsh:2007}).}
\label{fig:LateralPressure}
\end{figure*}

\color{black} For the sake of illustration, in the sequel we focus on the active receptors, namely the RL complex, as reported in \cite{Pollaci:2016}. \color{black}  In Figure \ref{fig:LateralPressure} it is shown a schematic of a TM domain involved in the conformational changes and the lateral pressure RL profile $\hat{\pi}(\textbf{x}, z, t)$ (marked with the green line in Figure \ref{fig:LateralPressure}.b). 
Indeed, this lateral pressure exhibits higher magnitudes in the zones where the lipid head groups interact with the top and bottom sites of the TMs. Therefore  $\hat{\pi}(\textbf{x}, z, t)$ performs work against the lateral variation of area, namely the local area change, $\hat{J}(\textbf{x}, z, t)$ at each level $z$ across the thickness.
%%%%%%%%%%%%%%%%%%%%%%%%%%%%%%%%%%%%%%%%%%%
We approximate the lateral pressure profile $\hat{\pi}(\textbf{x},z,t)=\hat{\pi}(\textbf{x},-z,t)$ with an even piecewise function \color{black} as displayed in red in Figure \ref{fig:LateralPressure}.b) and defined as follows:\color{black}
{\small
\renewcommand\arraystretch{1.5}
\begin{equation}
\label{eq:laterl_pressure_profile}
\hat{\pi}(\textbf{x},z,t) = \left\{
\begin{array}{ll}
-p & 0< z <\dfrac{h}{2} -\phi\\ 
\dfrac{q}{\phi/2} \left[ z - \left(\dfrac{h}{2} - \phi\right)\right] & \dfrac{h}{2} - \phi < z < \dfrac{h}{2} - \dfrac{\phi}{2}\\ 
q - \dfrac{q}{\phi/2} \left[ z - \left(\dfrac{h}{2} - \dfrac{\phi}{2} \right)\right]& \dfrac{h}{2} - \dfrac{\phi}{2}  < z < \dfrac{h}{2}
\end{array} 
\right.
\end{equation}
}
Here $h(\textbf{x},t)$ is the current thickness of the membrane at $\textbf{x}$ at the time $t$, $q$ denotes the value of repulsive pressure arising because of the contrast of the lipid headgroup against the receptor domain, and $p$ is the value of the attractive pressure along the hydrocarbon chain region arising to produce a self-balanced pressure profile. 
%
%=========================================
% Work done by lateral pressure
%=========================================
%\subsection{Work done by lateral pressure}
Balance through the thickness implies that the following relationship must hold:
\begin{equation*}
2 p \left(\dfrac{h}{2}- \phi \right) = 4 \, \dfrac{q \, \phi/2}{2},
\end{equation*}
hence $p$ and $q$ turn out to relate as follows:
\begin{equation}
\label{p_q}
p =  \frac{q}{2} \frac{\phi}{h - \phi} .
\end{equation}
The work $\tilde{w}_{\xi}$ done by the lateral pressure \eqref{eq:laterl_pressure_profile} against the change in lateral stretch $\hat{J}(\textbf{x},z,t)$ can be expressed as:
\begin{equation}
\label{W}
\tilde{w}_{\xi} = \pi\, d_s \,h \, \hat{w}_{\xi} ,
\end{equation}

where $\hat{w}_\xi$ denotes the work done per unit area and $\pi\, d_s \,h$ is the exhibited lateral surface of the TM. 

If we set
\begin{equation*}
\theta = z - \left( \frac{h}{2} - \phi\right) \qquad \text{and} \qquad \theta' = z - \left( \frac{h}{2} - \frac{\phi}{2} \right), 
\end{equation*}
the quantity $\hat{w}_\xi$  can be evaluated as follows:
{\small
\begin{equation}
\label{W*}
\begin{aligned}
\hat{w}_\xi & = 2 \int_0^{\frac{h}{2}} \hat{\pi}(\textbf{x},\,z,t) \,   \hat{J}(\textbf{x},\,z,t)  \, dz \\
&
\begin{aligned}
= 2 \Bigg( -p \int_{0}^{h/2 - \phi}  \hat{J}(\textbf{x},\,z,t)  \,d z  & + q\frac{1}{\phi/2} \int_{0}^{\frac{\phi}{2}} \theta \, \hat{J}(\textbf{x},\,\theta+\frac{h}{2}-\phi,t) \, d\theta
+  q \frac{\phi}{2}  +\\
& - q\frac{4}{\phi}  \int_0^{\frac{\phi}{4}} \theta' \hat{J}(\textbf{x},\, \theta'+\frac{h}{2}-\frac{\phi}{2},t) \, d \theta'
\end{aligned}
\\
&\approx 2 \left[ 
q \frac{\phi}{2}  
- p \int_{0}^{\frac{h}{2} - \phi}  \left[ \hat{J}(\textbf{x},\,0,t) + z \, \hat{J}_{z}(\textbf{x},\,0,t)   \right] \,d z 
\right] \\
&= 2 \left[q \frac{\phi}{2}  - p \left( \frac{h}{2} - \phi\right) \,J \right]  \\
&= \, \frac{\gamma_{RL}}{d_s} \, \phi \left(1 - J \right)
\end{aligned}
\end{equation}
}
\noindent \color{black} The result arises \color{black} after utilizing \eqref{p_q} relating $p$ to $q$
and upon recognizing that the surface tension $\gamma_{RL}$ of the RL compound at the lipid headgroup- is related to the lateral pressure as 
\begin{equation*}
q = \gamma_{{RL}}/d_s
\end{equation*}
where $d_s$ denotes the diameter of a TM domain.

Bearing in mind that $h=h_0/\lambda$, where $\lambda=J$, an approximaion for the work done by the lateral pressure is the computed through \eqref{W}, namely $\tilde{w}_{\xi} = \pi\, d_s \,h \, \hat{w}_{\xi}$, by virtue of \eqref{W*}, i.e.
\begin{equation}
\begin{aligned}
\tilde{w}_\xi & = \pi \, \phi \, h_0 \, \gamma_{RL} \, \left(\frac{1}{\lambda} \, - 1\right). \label{Work_lateral_pressure}
\end{aligned}
\end{equation}

% Because of relation \eqref{W}, namely $w_{l} = \pi\, d_s \,h \, \hat{w}_{l}$, the work per unit area and per unit receptor reduced with respect to the reference thickness reads as follows:

% \begin{equation}
% \hat{w}^0_l=\frac{\phi}{ds}\gamma_{RL}\left(\frac{1}{\lambda} \, - 1\right)=\omega\, \left(\frac{1}{\lambda} \, - 1\right)
% \end{equation}
With the identification 
\begin{equation}\label{omega-xi}
\omega_\xi =  \pi \, \phi \, h_0 \, \gamma_{RL}
\end{equation}
equation \eqref{Work_lateral_pressure} yields
\begin{equation}
\begin{aligned}
\tilde{w}_\xi & = \omega_\xi \, \left(\frac{1}{\lambda} \, - 1\right). \label{Work_lateral_pressure-1}
\end{aligned}
\end{equation}
namely \textit{the} (opposite of the) \textit{work per unit area and per unit receptor, reperesents the contribution to the chemical potential $\mu_\xi$ in \eqref{Pmudef2} due to mechanical actions associated with conformational changes, against which the lateral pressure performs work}. 

The remaining terms in \eqref{ChemPot}, namely the entropic one, $\eta_\xi \, \log (\xi / \xi^0)$, and the reference value $\tilde{\mu}_\xi^0$ add together and sum up to $\tilde{w}_\xi$ to deliver the complete expression of the chemical potential for the RL compound. Obviously, for the MRPs, namely the transporters, the reasoning is analog, thereby retrieving relation \eqref{Pmudef2} for $n_2=\zeta$, i.e. the analog expression for the chemical potential for such species. %$\mu_\zeta$
 
% Once ${w}_l$ gets then multiplied by the density of active receptors in order to give the chemo-mechanical coupling term derived in the potential \eqref{Wnidef} by means of thermodynamic considerations at the membrane macroscale. In other words, besides the entropic term depending on the species density $\xi$

The chemical stress \eqref{ChemStress} can be revisited in the light of \eqref{Work_lateral_pressure-1}. In particular, for the RL species $n_1= \xi$ it is worth noting that $\partial \tilde{\mu}_\xi / \partial \lambda= - \partial \tilde{w}_\xi / \partial \lambda$ and hence the chemical stress associated with the RL compound reads as follows:
\begin{equation}\label{ChemStress-xi}
\tilde{P}_\xi = - (\xi^0-\xi) \, \frac{\partial \tilde{w}_\xi} {\partial \lambda}.
\end{equation}
It is then worth emphasizing that, \textit{in spite of the fact that the lateral stress profile between TMs and lipids is self-balanced across the thickness, the rate of change of its work} (done against the lateral variation of area) \textit{times the relative density of the RL compound actually produces a quota of the membrane stress}. Of course for the MRPs an analog reasoning follows.
\section*{Appendix C} \label{SB2}
\subsection*{Conformational changes, remodelling and associated energetics}
\renewcommand{\theequation}{C.\arabic{equation}}
  % redefine the command that creates the equation no.
\setcounter{equation}{0}  % reset counter
%%%%%%%%%%%%%%%%%%%%%%%%%%%%%%%%%%%%%%%%%%%

\begin{figure}[htb!]
\centering
\includegraphics[width=0.95\columnwidth]{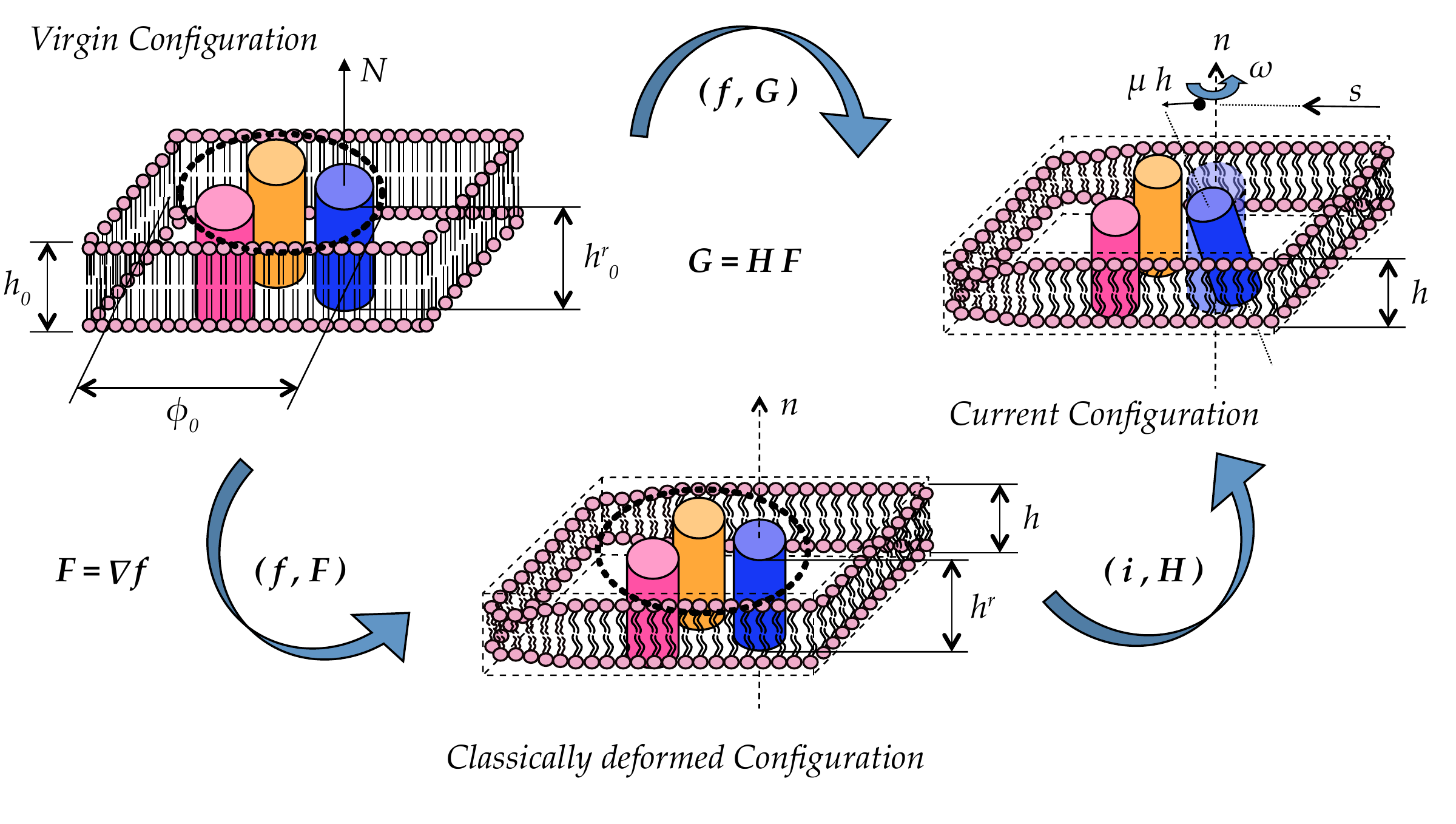}
\caption{Conformational changes and membrane deformation.}
\label{fig:StructuredDeformations}
\end{figure}

A schematic of the kinematics of the cell membrane in the presence of \textit{conformational changes} and, possibly, \textit{remodelling} of the lipid membrane is displayed in the sequel in Fig.\ref{fig:StructuredDeformations}. For the sake of illustration, this focuses on active receptors alone, although the treatment for active transporters (MRPs) can be done in the same way. 
Geometrical changes occuring among the TransMembrane domains of TM3, 5 and 6 (see \ref{fig_cell_with_receptors_and_transporters}) are shown together with the deformation of the lipid bilayer. 
As pointed out in Sect. 3.2, the
submacroscopic changes caused by the formation of the RL compound (i.e. the receptor-ligand binding) do cause a remodelling of the
cell membrane. This is due to a change of the density of active receptors thereby determining the re-organization of the surrounding
lipids.

\begin{figure}[htb]
\centering
\includegraphics[width=0.7\columnwidth]{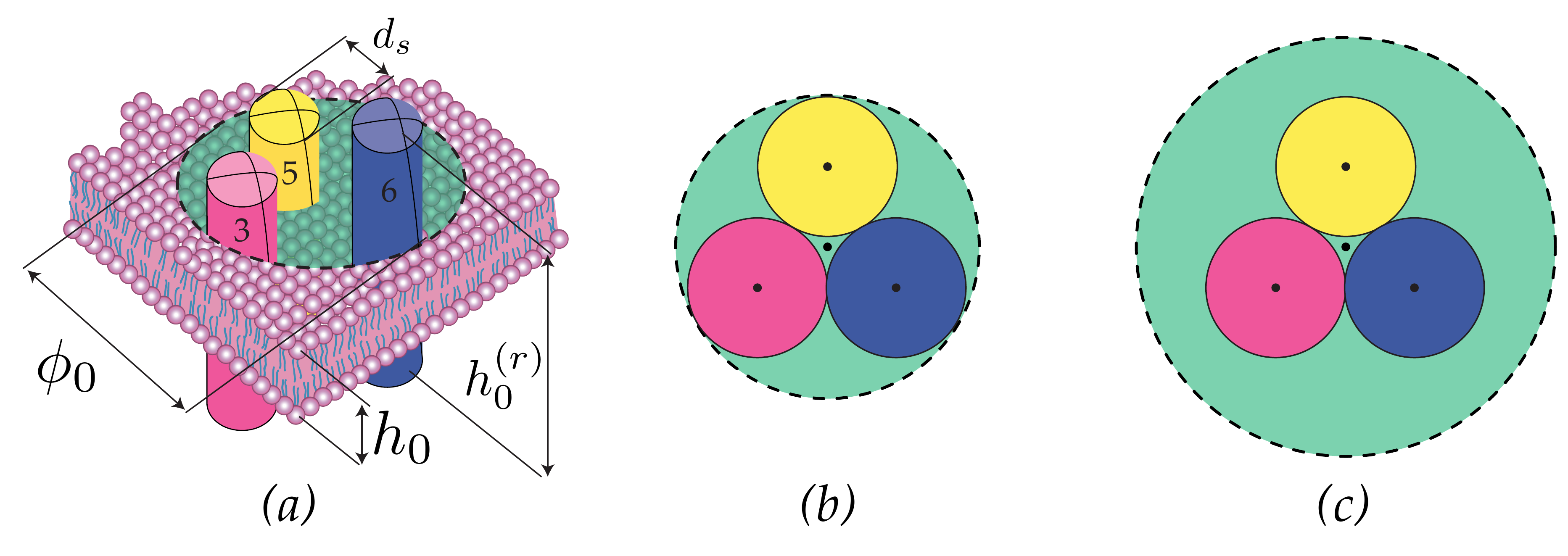}
\caption{Room space for conformational changes in active receptors as in \cite{Pollaci:2016}: (a) schematic of TM3, 5 and 6, (b) minimum and regular room space among the domains involved in the conformational changes.}
\label{fig_room_space}
\end{figure}

\textit{Structured Deformations} (see e.g. \cite{deseri:2003}, \cite{Deseri:2010}, \cite{deseri:2012}, \cite{deseri:2015}, \cite{deseri:2019}, \cite{palumbo:2018}) is a multiscale geometric framework that permits to account for both conformational changes and remodelling arising at the submacroscopic level. Such a framework does allow for envisioning three full configurations for the membrane.
%are shown to interpret the geometrical changes of the cell membrane due to both
%
In particular, on the top-left side of  Fig.\ref{fig:StructuredDeformations}, a schematic of a piece of the membrane in its virgin configuration is sketched. 
%For the sake of illustration, only the TMs involved in the conformational changes of a single GPCR are drawn. 
%
Besides the reference thickness for the membrane and for the TMs, $h_0$ and $h_0^r$  respectively, and their counterparts in the current configurations, $h$ and $h'$, the quantities highlighted there are the normal $\textbf{N}$ to the mid-plane of the lipid membrane in the virgin configuration,  its counterpart $\textbf{n}$ in the deformed configuration, the reference value $\phi_{\sss{T}}$ of the diameter of each TM,  and the available diameter $\phi_0$ for such movements to arise. The quantity $\rho_{\sss{T}}:=\phi_0/\phi_{\sss{T}}$ denotes the \textit{room space} available between the TMs for conformational changes in the virgin configuration, this indicates the \textit{degree of packing} of the TMs. The minimum value of the \textit{room space} $\rho_{\sss{T}}$ is estimated to be $1+2/\sqrt{3}$ (closest packing of the three domains involved in the movements), hence its value is taken to be between such a value and $3$, corresponding to the available room space of a TM in the middle of TM3, 5 and 6, as shown in the sequel in Fig.\ref{fig_room_space}. 
It is worth notning that this schematic allows for distinguishing two different deformed configurations: the classically deformed one and the current configuration. The latter is represented on the top right side of Fig. \ref{fig:StructuredDeformations}. Mathematically, this is described through a given Structured Deformation, namely a pair $(\textbf{y}, \textbf{G})$, where $\textbf{y}$ is the deformation mapping from the virgin configuration, and $\textbf{G}$ is a tensor field which would be equal to the gradient of $\textbf{y}$ if no conformational changes would occur. The lower part of  Fig. \ref{fig:StructuredDeformations} offers an explanation of such interpretation through the multiplicative decompostition $(\textbf{y},\textbf{G}) \,=\, (\textbf{i},\textbf{H}) \circ (\textbf{y},\textbf{F})$ of the pair $(\textbf{y},\textbf{G})$  introduced above. An intermediate classically deformed configuration is represented in the bottom of this figure where $(\textbf{y}, \textbf{F})$, with $\textbf{F} = \text{Grad}$ $\textbf{y}$, maps points in the virgin configuration in this intermediate one. This classical deformation is followed by a non-classical one. The latter is represented by a pair $(\textbf{i}, \textbf{H})$ and it is displayed in Fig. \ref{fig:StructuredDeformations}: here $\textbf{i}$ is the identity mapping (leaving then the macroscopic configuration untouched), while $\textbf{H}=\textbf{G}\textbf{F}^{-1}$   is a tensor field accounting for all the local conformational changes.

It is also worth noting that this representation is fully compatible with the schematic depicted in Fig.\ref{fig:Stress0} in the main text.  As pointed  out in \cite{deseri:2003}, a factorization of the structured deformation $(\textbf{y},\textbf{G}) \,=\, (\textbf{y},\textbf{F})  \circ (\textbf{i},\textbf{K}) $ such that a classical deformation follows a non-classical one where the macroscopic deformation of the body does not change, while conformational changes such that $\textbf{K}=\textbf{F}^{-1}\textbf{G}$ can occur at the submacroscopic level. Obviously there is a one-to-one relationship between the configurational changes described  (through $\textbf{H}$) from the classically deformed configuration to the current configuration and their description starting from the virgin configuration (through $\textbf{K}$). Indeed, pure geometry is suggestive of the fact that the deformation of the membrane plays a key role (through $\textbf{F}$ $=$ Grad $\textbf{y}$)  on carrying over the information related to the conformational changes, as  $\textbf{K}=\textbf{F}^{-1} \textbf{H} \textbf{F}$.

It is worth noting that the definition of the remodelling factor introduced in \eqref{RemodelingTerm} can be exactly written as
\begin{equation}
K_r \, =\, det \textbf{K},
\label{remodelling-Structured Deformation}
\end{equation}
as the change in volume between the virgin configuration and the reference configuration is solely due to disarrangements which, in this case, are due to (submacroscopic) remodelling.

Furthermore, it is obvious that $det \textbf{K} \, = \, det \textbf{H}$. Henceforth, the evaluation of the remodelling factor can either be done upon following the scheme displayed in Fig.\ref{fig:Stress0} ($(\textbf{y},\textbf{F})  \circ (\textbf{i},\textbf{K})$, namely by letting disarrangements to act first, then followed by a classical deformation) or the one in Fig.\ref{fig:StructuredDeformations} ($(\textbf{i},\textbf{H}) \circ (\textbf{y},\textbf{F})$, i.e.  a classical deformation first, then the conformational changes).

\subsection*{Conformational energy: the case of active receptors}
\label{sectionB2}
As is has been recalled in Fig.\ref{fig_cell_with_receptors_and_transporters}, conformational changes in active receptors are  mainly characterized by a rotation $\omega$ of TM6 about its axis and by a translation $\mu h$ of the TM6 domain with respect to TM3 and TM5 (see e.g. \cite{Ghanouni:2001}). 
%Here $h$ is the value of the membrane thickness in the current configuration and $\mu$ is the corresponding shear. 

Henceforth, the change in Helmholtz free energy per unit area and per unit receptor relative to the classically deformed configuration (as displayed in Fig. \ref{fig:StructuredDeformations}) due to such movements can be then interpreted as a result of two contributions, both arising in the current configuration (see \cite{Pollaci:2016}). For both of such terms, the change in entropy per unit area and per unit receptor is evaluated as in \cite{Pollaci:2016}. The change of entropy due to rotation is accounted for by generalizing the result obtained in \cite{Finkelstein:1989}, namely
\begin{equation}
\label{eq:rotational}
\varphi_{\scriptscriptstyle CR}^{(1)} = A \log \left(\frac{\omega}{2 \pi^{2/3}} \right) ,
\end{equation}
where $\omega$ represents the rotation of the TM domains involved in the conformational change about the normal $\textbf{e}_3$ to the mid-plane of the membrane in the current configuration, $A$ is a normalization constant and $A \cdot K_B T$ represents the conformational energy level per unit receptor per unit area at the reference angle $\omega^* = 2 \pi^{2/3} e$.
The local translation measure is relative to the free volume available for the conformational changes in the current configuration. This is calculated by multiplying the current value of the thickness $h$ of the cell membrane by the available area  $\pi (d_0/2)^2  J$, where $d_0$  represents the referential  diameter of the zone in which conformational changes of the TMs can occur. It is worth recalling that $J=h_0/h$ depends on the location $\textbf{x}=(x_1, x_2)$ of the mid plane of the bilayer. As recalled in Sect. \ref{sectionB3},  it has been shown  that $J$ is the order parameter in lipid membranes (see e.g. \cite{Zurlo:2006, Deseri:2008, Deseri:2013}), allowing for discriminating whether or not lipids exhibit ordered ($J=1$ indicates the completely straight configuration of the lipid tails) or disordered phases ($J>1$, curlier configuration). Henceforth, the resulting change in entropy due to translational changes may be written as follows:
\begin{equation}
\label{eq:translational}
\varphi_{\scriptscriptstyle CR}^{(2)} = A \log \left(\frac{\mu \, h}{ \left(J \, (d_0/2)^2  \, h \right)^{1/3}} \right),
\end{equation}
where $\mu \,h$ represents a measure of the local translation of the TMs with respect to current location  and $\left(J \,(d_0/2)^2 \, h\right)^{1/3}=\left((d_0/2)^2 \, h_0\right)^{1/3}$ is a reference measure of such translation.  Upon introducing the  ``room space" parameter 
\begin{equation}
\rho_T=d_0/d_s,
\label{room_space}
\end{equation}
which measures how much room is available for the conformational changes of the TMs, as $d_s$ indicates the diameter of any of the TMs involved in such changes. 

The total \textit{conformational energy density per unit area and per unit (distribution density of) receptors} is obtained as the sum of both contributions \eqref{eq:rotational} and \eqref{eq:translational} above, i.e.
\begin{equation}
\label{eq:ConformationEnergy}
\varphi_{\scriptscriptstyle CR} = \varphi_{\scriptscriptstyle CR}^{(1)} + \varphi_{\scriptscriptstyle CR}^{(2)} = A \log\left(\kappa \frac{\eta}{J}\right) ,
\end{equation}
where 
\begin{equation}
\eta = \omega \, \mu
\label{eta}
\end{equation}
represents the \textit{conformational field} and $\kappa$ is a dimensionless geometric constant defined as follows:
\begin{equation}
\label{eq:K}
\kappa = \frac{1}{2^{}} \left( \frac{2\,h_0}{\pi \, d_s \, \rho_{\sss{T}}} \right)^{\frac{2}{3}} .
\end{equation}
Accordingly to equation \eqref{eq:K} and a range of geometric quantities collected from literature in \Tab{tabF:K_parameters}, a range for $\kappa$ can be then determined to be $\kappa_{\min} \leq \kappa \leq \kappa_{\max}$, where $\kappa_{\min} = 0.587$, $\kappa_{\max} = 1.634$ and $\kappa_m = (\kappa_{\min}+\kappa_{\max})/2 = 1.11$ will be used.

\begin{table}[htbp]
\centering
\begin{tabular}{ccc}
\toprule
\small{thickness $h_0$} & \small{diameter $d_s$} & \small{room space $\rho_{\sss{T}}$}  \\
$[nm]$ & $[nm]$  & - \\
\midrule 
$3\div 6$ & $0.3\div 0.5$ & $2.15 \div 3$ \\
\bottomrule
\end{tabular}
\caption{\textbf{Table}  \ref{tabF:K_parameters}. Ranges of values of geometrical parameters assumed from  \cite{Lee:2004, Hiden:2007, Pollaci:2016}, while the room space $\rho_{\sss{T}}$ is estimated from \Fig{fig_room_space}. \label{tabF:K_parameters}} 
\end{table}
\noindent In \cite{Pollaci:2016} bounds for $\eta$ are found in in terms of the ratio $h/h_0$. Nonetheless, for a given set of geometrical parameters $h_0, d_s, \rho_T$, \color{black} it can be shown that $\eta^*=K^{-1}$ represents the lowest possible value for $\eta$. \color{black}  Maximum admissible values of the conformational changes $\eta$ can also be estimated by considering extreme configurations of a single TM %, as shown in \Fig{fig:TM_positions}%
. Indeed $\eta$ is composed by a rotation, $\omega$ in the range $(0, \pi]$ (i.e. the value $0$ cannot be achieved), and a shear, $\mu = \tan \, \alpha$  where $\alpha$ is an angle across thickness (see \Fig{fig:StructuredDeformations}). The angle $\alpha$ can conceivably achieve values up to $\pi/4$, due to the presence of the surrounding TMs. Then, the conformational field is found in the following interval: $\eta^* \, \leq \, \eta \, \leq \, \pi$. 

\noindent The result \eqref{eq:ConformationEnergy} is a generalization of  \cite{Finkelstein:1989} and \cite{Murray:2002}, as the entropic changes are measured while accounting for the deformation of the lipid membrane.

\subsection*{Conformational energy and work done by the lateral pressure: the case of active receptors}
\label{sectionB3}
The contribution to the total Helmholtz free energy $W$ 
of the energetics of active receptors, RL, and transporters, MRPs,
has been worked out in Sect. \ref{sec:3M}. In particular, equation \eqref{Wnidef} for the RL compounds specializes as follows
\begin{equation}\label{Wnidef-xi}
W_{\xi}(\lambda^{-1},\xi)= \eta_\xi \, \xi \, \log \frac{\xi}{\xi^0}+(\xi-\xi^0)\left(\omega_\xi\left(\lambda^{-1}-1\right) -\tilde{\mu}_{\xi}^0 -\eta_\xi \right).
\end{equation}
Equation \eqref{Pmudef2} particularized for such a species and integrated over the domain $\Omega_a$ in the intermediate configuration, can be considered. Its variational derivative  with respect to $\xi$, the density of RL, leads the following expression for the chemical potential 
\begin{equation}
\tilde{\mu}_\xi = \tilde{\mu}_\xi^0 - \omega_\xi \, \left(\lambda^{-1}-1\right) -\eta_\xi \, \log \frac{\xi}{\xi^0} .\label{Pmudef2-2}
\end{equation}
In \hyperref[{lateralpress}]{Appendix C} it has been obtained that the term $\omega_\xi \, \left(\lambda^{-1}-1\right)$ appearing both in \eqref{Wnidef-xi} and \eqref{Pmudef2-2} solely regards the work done by the lateral pressure against the local area variation arising across the membrane during conformational changes. 
%By looking at the latter relation, namely the one giving the chemical potential for active receptors, 
%

\noindent Mixing of the the RL (active receptors) throughout the membrane contributes to lower the energy of the system in a purely entropic way (see e.g. \cite{Rangamani:2014}). Upon identifying 
\begin{equation}
\eta_\xi = K_B \, T,
\label{etacsi}
\end{equation}
 where $K_B$ is the Boltzmann constant, $T$ is the absolute temperature of the bath in which the cells are embedded,  i.e.
\begin{equation}
\label{eq:ligand-receptor-diffusion}
K_B T ~ \xi \left(-e_{RL} + \log \left(\frac{\xi}{\xi^0} \right) \right),
\end{equation}
where $e_{RL}$ is the specific activation energy for the complex receptor-ligand. %and $\xi_0$ denotes an average density of receptors computed by assuming that all them are active and uniformly distrubuted across the surface of the membrane. For the sake of simplicity, this quantity is named \textit{density of activable receptors}.

The corresponding energy density \eqref{eq:ConformationEnergy} per unit receptors, per unit area and per unit $K_B T$ due to conformational changes and ideal mixing allows for writing the quota of the total Helmholtz free energy associated with $\xi$ as follows:
% summing up all the contributions, i.e relationships  \eqref{eq:ligand-receptor-diffusion} \eqref{eq:cAMP-transporter-diffusion} \eqref{eq:ConformationEnergy} \eqref{eq:psi-star}, the Helmholtz free energy of the system is obtained in the following form:
%
\begin{equation}
\label{eq:energy-RL}
\begin{aligned}
\int_{\Omega_a} \xi &\left[-e_{\scriptscriptstyle RL} + \log\left(\frac{\xi}{\xi^0} \right) - \varphi_{\scriptscriptstyle CR} \right]\, d\Omega_a. \\
\end{aligned}
\end{equation}

The variational derivative of the overall energy with respect to $\xi$ yields the following expression of the corresponding chemical potential 
\begin{equation}
\label{eq:chemical_potential_RL}
\tilde{\mu}_\xi = K_B T \left( -e_{RL} + 1 + \ln \left( \frac{\xi}{\xi^0} \right) - \left(\xi  \,\varphi_{\scriptscriptstyle CR} \right)_{,\xi}\right).
\end{equation}

As in \cite{Pollaci:2016}, by taking \eqref{etacsi} into account, the comparison between \eqref{eq:chemical_potential_RL} and \eqref{Pmudef2-2}  establishes the consistency of both expressions for the chemical potetial of the RL compound and leads to the following relation among the density of active receptors $\xi$, the conformational field $\eta$ and the membrane thickness change $h$:
\begin{equation}
\label{eq:X-1}
\left(\frac{\xi}{\xi^0} \right)^{-1} = 1 - \dfrac{C}{\dfrac{h}{h_0} -1 } \ln \left(K \dfrac{h}{h_0} \eta \right).
\end{equation}
where
\begin{equation}
\label{eq:C}
C = \dfrac{K_B T A}{\pi \phi_T h_0 \gamma_{RL}}.
\end{equation}
\begin{figure}[htbp]
\centering
\includegraphics[width=0.5\columnwidth]{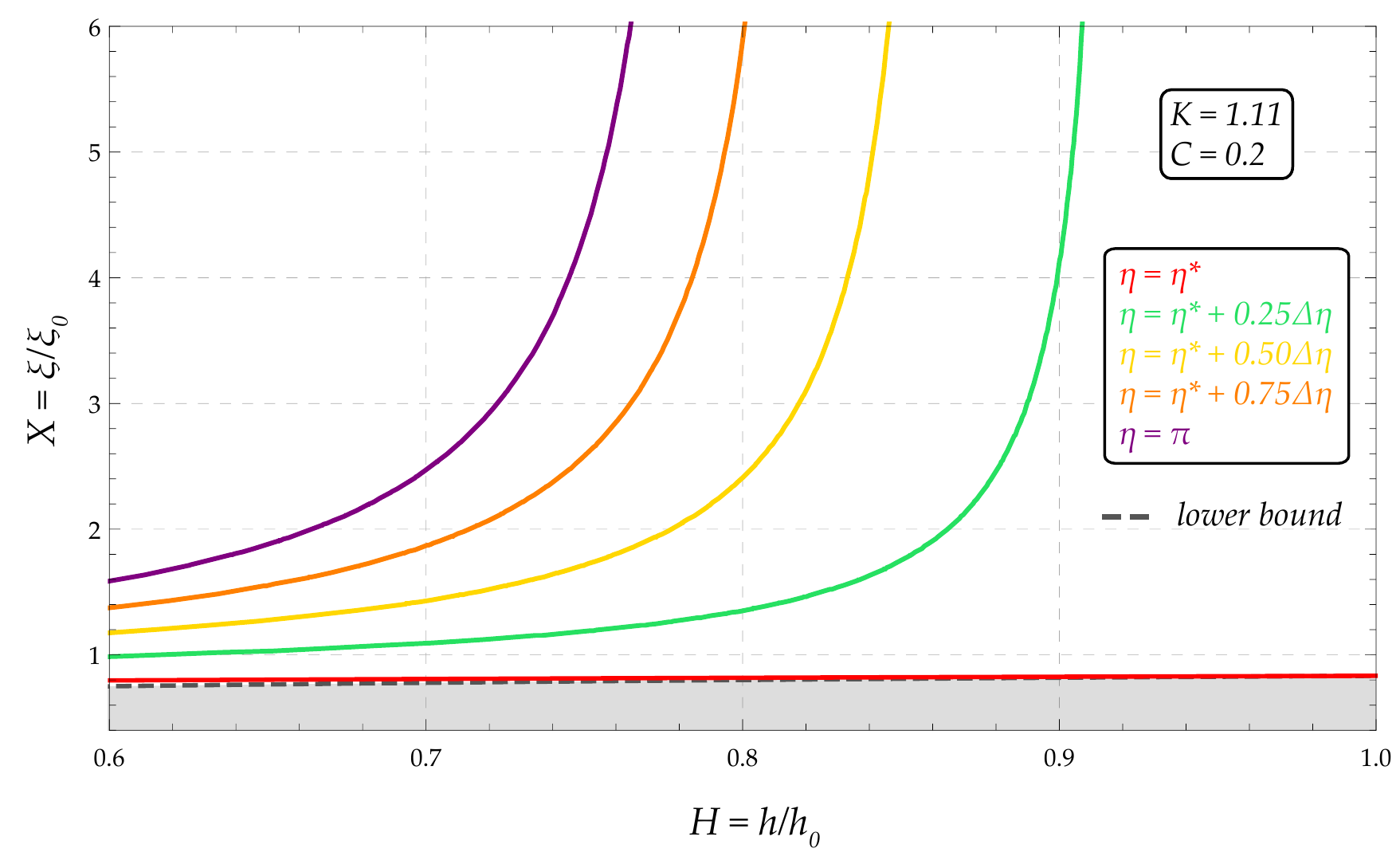}
\caption{Behavior of $\xi/\xi^0$ vs $h/h_0$ using $\eta$ as parameter. Here $\eta^* = 1.07$ represents a basal value and $\Delta \eta = 2.06$ (see \cite{Pollaci:2016}).}
\label{fig:raft_conformation}
\end{figure}
In \Fig{fig:raft_conformation}, slices obtained by fixing values of $\eta$ are shown, namely the quantity $X = \xi/\xi^0$ as a function of the relative thickness change $H = \lambda^{-1}= h/h_0$. This shows that relative thickness change $H$ increases with an increase of active receptor density $X$. This result, is found to be in agreement with the experimental findings of Lefkovitz \cite{Lefkowitz:2007} and Kobilka \cite{Kobilka:2007}. As in a purification process the addition of cholesterol to the membrane induces lipid ordering, this actually enhances the clustering of activated receptors on rafts. Our model is then consistent to this observation already at the constitutive level. Furthermore, in \cite{Pollaci:2016} it is shown that equation \eqref{eq:X-1} allows one to obtain the conformation in terms of relative distribution of active receptors $X$ and of the thinning field $H$, in the following form: 

\begin{equation}
\label{eq:relationship_eta}
\eta = \frac{1}{H K} \exp \left( - \frac{(1-H)(1-X)}{C \, X}\right) . 
\end{equation}
The relative receptor density $X$ is a positive definite quantity  and it is increasing with $H$. In \cite{Pollaci:2016} it is shown that the simultaneous occurrence of both these conditions implies the occurrence of the following inequalities:
\begin{subequations}
\label{eq:eta_bounds}
\begin{align}
\eta < \frac{1}{K~H} \exp \left(\frac{1-H}{C} \right) = \eta_{\sss{UB}}, \\
\eta \ge \frac{1}{K~H} \exp \left(\frac{H-1}{H} \right) = \eta_{\sss{LB}},
\end{align}
\end{subequations}
for given values of $H$, $C$ and $K$. Further investigations (see Table \ref{tab:A}  below) allow for estimating $C$ in the range $[0.05 \div 0.75]$, while the value taken for this parametric study is $C = 0.3$. 

%
%\begin{figure}[htbp]
%\centering
%\includegraphics[width=0.9\columnwidth]
%{figures/figF_raft_kobilka}
%\caption{Schematic of raft in a membrane: A) cytosol, B) extracellular space, 1) non-raft membrane, 2) lipid raft, 3) lipid raft associated transmembrane protein, 4) non-raft membrane protein, 5) glycosylation modification, 6) GPI-anchored protein, 7) cholesterol, 8) glycolipid. Reproduction from \footnotesize{http://cellbiology.med.unsw.edu.au/units/science/lecture0803.html}.}
%From \scriptsize{http://cellbiology.med.unsw.edu.au/units/science/lecture0803.html}
%\label{fig:Kobilka-rafts}
%\end{figure}

%
\begin{table}[htbp]
\caption{\textbf{Table} \ref{tab:A} Estimated values for $C$. For each choice of parameters, the first pair of rows is referred to $A =1$, while the second refers to $A = 5$. Values of $\gamma_{\sss{RL}}$ come from literature \cite{Ollila:2010, Marsh:2007, Pollaci:2016}.} \label{tab:A}
\centering
\small
\begin{tabular}{ccccc}
\toprule
 $h_0 \, [nm]$  & $\phi_{\sss{T}} \, [nm]$ & $\gamma_{\sss{RL}} \, [mN/m]$ & $A/C$ & $C$ \\
\midrule 
\multirow{2}{*}{3} & \multirow{2}{*}{0.5} & \multirow{2}{*}{10} & \multirow{2}{*}{11.01} & 0.091\\
 &  & & & 0.454\\
\midrule 
\multirow{2}{*}{6} & \multirow{2}{*}{0.3} & \multirow{2}{*}{10} & \multirow{2}{*}{13.21} & 0.076\\
 &  & & & 0.379\\
\midrule 
\multirow{2}{*}{3} & \multirow{2}{*}{0.5} & \multirow{2}{*}{35} & \multirow{2}{*}{38.52} & 0.026\\
 &  & & & 0.129\\
 \midrule 
\multirow{2}{*}{6} & \multirow{2}{*}{0.3} & \multirow{2}{*}{35} & \multirow{2}{*}{42.22} & 0.022\\
 &  & & & 0.108\\
\bottomrule
\end{tabular}
\end{table}

%%%%%%%%%%%%%%%%%%%%%%%%%%%%%%%%%%%%%%%%%%%%%%%%%%%%%%%%%%%
\newpage
%%%%%%%%%%%%%%%%%%%%%%%%%%%%%%%%%%%%%%%%%%%%%%%%%%%%%%%%%%%
\bibliographystyle{unsrt}  

\end{document}

%% file: definitions.tex
\usepackage{isomath}
\usepackage{amsmath}
\usepackage{amssymb}
\usepackage{amscd}
\usepackage{amsfonts}

\newcommand{\beq}{\begin{equation}}
\newcommand{\eeq}{\end{equation}}
\newcommand{\beqs}{\begin{eqnarray}}
\newcommand{\eeqs}{\end{eqnarray}}

\newcommand{\grad}{\mathop{\rm grad}\nolimits}